\documentclass[fleqn,usenatbib]{mnras}

\usepackage{newtxtext,newtxmath}

\usepackage[T1]{fontenc}

\DeclareRobustCommand{\VAN}[3]{#2}
\let\VANthebibliography\thebibliography
\def\thebibliography{\DeclareRobustCommand{\VAN}[3]{##3}\VANthebibliography}

\usepackage{graphicx}	
\usepackage{amsmath}	
\usepackage[section]{placeins}
\usepackage[english]{babel}
\usepackage{lineno}
\PassOptionsToPackage{hyphens}{url}
\usepackage{ulem}
\usepackage{hyperref}
\usepackage[dvipsnames]{xcolor}
\usepackage{multirow}

\newcommand{\hinvMsun}{h^{-1} {M_\odot}} 
\newcommand{\hinvhinvLsun}{h^{-2} {L_\odot}} 
\newcommand{\Mh}{M_{\rm h}}                  
\newcommand{\logMh}{\log (M_{\rm h}/[h^{-1}{M_\odot}])}  
\newcommand{\logMg}{\log (M_{\rm g}/[{h^{-1}M_\odot}])}  
\newcommand{\Yang}{\tt Yang~et~al.}          

\newcommand{\Mnl}{M_{\rm nl}}

\def\aap{A\&A}

\def\apj{ApJ}
\def\apjs{ApJS}
\def\apjl{ApJL}

\def\mnras{MNRAS}
\def\aj{AJ}

\def\araa{ARA\&A}

\title[CCMD groups VS. SDSS groups]{The conditional colour–magnitude distribution: II. A comparison of galaxy colour and luminosity distribution in galaxy groups}

\author[H. Xu et al.]{
Haojie Xu$^{1,2,3}$\thanks{E-mail: \href{mailto:haojie.xu@shao.ac.cn}{haojie.xu@shao.ac.cn} (HX)},
Zheng Zheng$^{4}$\thanks{E-mail: \href{mailto: zhengzheng@astro.utah.edu} {zhengzheng@astro.utah.edu} (ZZ)},
Xiaohu Yang$^{2,3,5}$,
Qingyang Li$^{2,3}$, 
and Hong Guo$^{1}$\\
\\
$^{1}$Shanghai Astronomical Observatory, Chinese Academy of Sciences, Nandan Road 80, Shanghai 200240, China\\
$^{2}$ Department of Astronomy, Shanghai Jiao Tong University, Shanghai 200240, China\\
$^{3}$ Key Laboratory for Particle Astrophysics and Cosmology (MOE)/Shanghai Key Laboratory for Particle Physics and Cosmology, China\\
$^{4}$ Department of Physics and Astronomy, University of Utah, 115 South 1400 East,
Salt Lake City, UT 84112, USA\\
$^{5}$Tsung-Dao Lee Institute, Shanghai Jiao Tong University, Shanghai 200240, China
}

\date{Accepted XXX. Received YYY; in original form ZZZ}

\pubyear{2023}

\begin{document}
\label{firstpage}
\pagerange{\pageref{firstpage}--\pageref{lastpage}}
\maketitle

\begin{abstract}
The Conditional Colour-Magnitude Distribution (CCMD) is a comprehensive formalism of the colour-magnitude-halo mass relation of galaxies. With joint modelling of a large sample of SDSS galaxies in fine bins of galaxy colour and luminosity, Xu et al. inferred parameters of a CCMD model that well reproduces colour- and luminosity-dependent abundance and clustering of present-day galaxies. In this work, we provide a test and investigation of the CCMD model by studying the colour and luminosity distribution of galaxies in galaxy groups. An apples-to-apples comparison of group galaxies is achieved by applying the same galaxy group finder to identify groups from the CCMD galaxy mocks and from the SDSS data, avoiding any systematic effect of group finding and mass assignment on the comparison.
We find an overall nice agreement in the conditional luminosity function (CLF), the conditional colour function (CCF), and the CCMD of galaxies in galaxy groups inferred from CCMD mock and SDSS data. We also discuss the subtle differences revealed by the comparison. In addition, using two external catalogues constructed to only include central galaxies with halo mass measured through weak lensing, we find that their colour-magnitude distribution shows two distinct and orthogonal components, in line with the prediction of the CCMD model. Our results suggest that the CCMD model provides a good description of halo mass dependent galaxy colour and luminosity distribution. The halo and CCMD mock catalogues are made publicly available to facilitate other investigations. 
\end{abstract}

\begin{keywords}
galaxies: distances and redshifts 
-- galaxies: haloes 
-- galaxies: statistics
\end{keywords}



\section{Introduction}
\label{sec:intro} 
The past two decades have seen enormous growth of studying the statistical relationship between the physical properties of galaxies and their dark matter halos (see \citealt{Wechsler2018} for a recent review), owing to large galaxy surveys. A through understanding of the galaxy-halo connection is yet to be achieved, as dark matter haloes cannot be directly observed.  A full comprehension of such a connection would not only provide us with insight into the complex physical processes of galaxy formation and evolution, but also enable us to infer the underlying dark matter distribution through galaxy surveys.

Among the physical properties of galaxies that can be directly observed through galaxy surveys, luminosity and colour are the two most indicative of star formation activity --- luminosity is a rough approximation of stellar mass, while colour encodes information on star formation rate and history \citep[e.g.][]{Blanton05b}. On the halo side, halo mass is the most important factor in the formation and growth of haloes and their clustering \citep[e.g.][]{Zentner2007, Tinker2011, Wang2018}. Therefore, understanding the relationship between galaxy luminosity, colour, and halo mass is the first step in establishing a comprehensive connection between galaxies and haloes.

The colour--magnitude--halo mass relation can be easily predicted by cosmological hydrodynamic simulations and semi-analytic galaxy formation models, which are computationally expensive and rely heavily on assumed subgrid physics or galaxy formation recipes. An alternative, empirical approach is through combining observed galaxy clustering with the theoretically known halo clustering. Galaxy clustering is observed to depend on both luminosity and colour \citep[e.g.][]{Norberg01, Norberg02,Zehavi02,Zehavi2005,Zehavi11, Coil08, Xu2016}, and halo clustering is dependent on halo mass \citep[e.g.][]{Mo96}. The colour--magnitude--halo mass relation can be established by tuning how galaxies with different luminosities and colours occupy haloes of different masses to reproduce the observed clustering. This is the core idea behind models such as the halo occupation distribution \citep[HOD;][]{Jing98, Seljak00, Berlind02, Zheng05} and the conditional luminosity function \citep[CLF;][]{Yang03}. The HOD/CLF framework has been widely used to model luminosity-dependent galaxy clustering \citep[e.g.][]{Yang03,vandenBosch03, Zehavi2005,Zehavi11}. 
Colour-dependent clustering can be modelled by parameterising the fraction of blue galaxies as a function of halo mass \citep[e.g.][]{Zehavi2005,vandenBosch03, Ross2009}. In \citet{Zehavi11}, it is found that varying the central-to-satellite ratio for galaxies in different colour subsamples is sufficient to capture the trend of colour-dependent clustering at fixed galaxy luminosity. There are also other approaches to modelling colour-dependent clustering \citep[e.g.][]{Skibba09}.

As an extension to the HOD/CLF framework, \citet{Xu2018} (the first paper in this series, Xu18 hereafter) proposes the conditional colour-magnitude distribution (CCMD) to link
galaxy luminosity and colour to dark matte halo mass.
CCMD deprojects the observed galaxy colour-magnitude diagram (CMD) into a halo-mass-dependent distribution with a decomposition into central and satellite galaxy populations. Motivated by the colour bimodality, each population is further divided into two components, pseudo-blue and pseudo-red\footnote{At ﬁxed luminosity, the bimodal colour distribution of galaxies can be well described by a superposition of two Gaussian components\citep[e.g.][]{Baldry04}. The two components can be broadly referred to as ‘red’ and ‘blue’ galaxies. However, the two components can substantially overlap each other. To avoid any possible confusion with the usual red and blue samples,  Xu18 named them pseudo-red and pseudo-blue components.}. At a ﬁxed halo mass, the magnitude and colour distribution of each pseudo-colour central population is described by a two-dimensional (2D) Gaussian distribution, while the luminosity distribution of the satellite population is parameterised as a Schechter-like function with colours following a luminosity-dependent Gaussian distribution.
The characteristic quantities that define the four components are allowed to vary with halo mass, motivated by previous work on HOD and CLF modelling \citep[e.g.][]{Yang08, vandenBosch13}.
The CCMD parameters are obtained by simultaneously fitting the galaxy abundances and projected two-point correlation function (2PCF) measurements of galaxy samples deﬁned by $\sim$80 ﬁne luminosity and colour bins, constructed from the Sloan Digital Sky Survey Data Release 7 Main Galaxy Sample \citep[SDSS DR7;][]{York00,Strauss02,DR7}.

The galaxy colour--magnitude--halo mass relation derived from CCMD modelling results can be compared to an array of galaxy-halo relations obtained from various methods. For example, Xu18 compared the central galaxy luminosity--halo mass relation based on the CCMD results with those from traditional HOD modelling \citep{Zehavi11}, galaxy-galaxy weak lensing \citep{Mandelbaum06}, satellite kinematics \citep{More11}, CLF modelling of galaxy clustering and galaxy-galaxy lensing \citep{Cacciato09}, and generally good agreements are found. As the galaxy-halo relation is supposed to be readily inferred from a galaxy group catalogue \citep[e.g.][]{Yang05b,Yang07,Yang2021,Tinker2021}, they also compared the CLF from an SDSS group catalogue and that from the best-fitting CCMD model and found differences. For example, compared to the CCMD model, the CLF of central galaxies from the SDSS group catalogues appears to be much narrower, and the CLF of satellite galaxies shows a much steeper drop at the luminous end.  However, this comparison turns out to be indirect, as the former is based on groups and the latter on haloes. Groups and haloes are not equivalent due to factors such as the group-ﬁnding algorithm, the purity and completeness of groups, the way halo mass being assigned, and the deﬁnitions of central and satellite galaxies \citep[][Campbell15]{Campbell15}. As suggested by Xu18, the best way to facilitate such comparisons is to construct mock galaxy catalogues from the CCMD modelling results and apply the same group-finding code as used in the observation to identify groups in the mock
\citep[e.g.][]{Reddick2013,Campbell15,Calderon2018}. 
This would enable direct and fair comparisons of the colour--magnitude--halo mass relation between the SDSS and CCMD groups.

In this paper, we present such a comparison. In addition to the CLF, we will compare the full distribution of galaxy luminosity and colour, including the division into contributions from central and satellite galaxies. This comparison will serve as a useful check for the CCMD model and will test the validity of the CCMD in reproducing the observed galaxy-group relation. Any revealed inconsistency may lead to a revisit of the assumptions in the CCMD model and further advance our understanding of the galaxy-halo relation.

The paper is structured as follows. In Section~\ref{sec:sdss_mock_gf}, we provide a brief overview of the observational data set used to constrain the CCMD parameters, the CCMD mock construction, and the group-finding algorithm. 
Section~\ref{sec:CCMDgroups_SDSSgroups} presents a comparison between the galaxy luminosity and colour distribution in the CCMD groups and in the SDSS groups. 
Section~\ref{sec:direct_ccmd} aims to uncover the CCMD features without relying on group mass or identified group membership, by binning the galaxy groups through dynamical mass and cross-matching with external central galaxy catalogues. 
We summarise the main conclusions in Section~\ref{sec:summary}.

Throughout the paper, all magnitudes and colours are in Petrosian magnitudes, and have been $K$-corrected to $z \sim 0.1$, the median
redshift of galaxies in the SDSS DR7 Main Galaxy Sample. The magnitude
is calculated by setting $h=1$, where $h$ is the Hubble constant in units
of 100${\rm km\, s^{-1}\, Mpc^{-1}}$. 
The term ``magnitude'' refers to ``absolute magnitude'', except when ``apparent magnitude'' is explicitly stated.
The base-10 logarithm is denoted as $\log$.
We employ a magnitude-dependent colour cut as in \citet{Zehavi11}, $(g - r)_{\rm cut} = 0.21 - 0.03M_r$, to separate blue and red galaxies.

\section{SDSS DR7, CCMD Mocks, and Group Finder}
\label{sec:sdss_mock_gf} 

In this work, we compare the galaxy contents in galaxy groups identified in SDSS data with those identified in mock galaxy catalogues based on the CCMD model. In Section~\ref{subsec:sdss}, we briefly introduce the SDSS data. The luminosity- and colour-dependent SDSS galaxy clustering measurements are used to constrain the CCMD model.
In Section~\ref{subsec:ccmdmock}, we describe the construction of CCMD mocks and
mock redshift surveys that mimic the observation.
To better understand how the group finder can be used to evaluate the colour--magnitude--halo mass relation,
a brief introduction of the group-finding algorithm is described in Section~\ref{subsec:gf}.

For simplicity, we adopt the following naming convention throughout the work:
\begin{itemize}
\item[$\bullet$]CCMD haloes -- $N$-body haloes used in constructing the CCMD mocks;

\item[$\bullet$]CCMD groups -- galaxy groups identified by applying the group finder to the CCMD mocks;

\item[$\bullet$]SDSS groups -- galaxy groups identified by applying the group finder to the SDSS DR7 data.
\end{itemize}
To differentiate the mass assigned to a galaxy group and that for a dark matter halo, we call the former group mass $M_g$ and the latter halo mass $M_h$.

\subsection{SDSS DR7}
\label{subsec:sdss}

We apply the group finder to the dataset SDSS DR7 \citep{DR7}. This is the dataset that the CCMD model is based on --- galaxy number densities and galaxy clustering measurements for galaxy samples in fine bins of luminosity and colour are simultaneously modelled with a global parametrisation of the CCMD to obtain constraints on the CCMM parameters (Xu18).  

Specifically, galaxy positions and properties are extracted from {\it bright1},
the large-scale structure sample of the NYU Value-Added Galaxy 
Catalog\footnote{\url{http://sdss.physics.nyu.edu/lss.html}} 
(NYU-VAGC; \citealt{Blanton05a}).
Before applying the group finder, we remove galaxies
that do not meet the following criteria: $0.01 \leq z < 0.2$ and $M_{r} < -18$.
The redshift cut is based on the redshift range of the widely used $\Yang$ group catalogue \footnote{\url{https://gax.sjtu.edu.cn/data/Group.html}}
(see Section~\ref{subsec:gf}).
The $r$-band absolute magnitude cut of $M_{r} < -18$ is due to the fact that the CCMD parameters are constrained by galaxy samples more luminous than -18 mag (Xu18).

\subsection{CCMD mock catalogues}

The group finder is applied to mock galaxy catalogues constructed based on populating galaxies into dark matter haloes from an $N$-body cosmological simulation, according to CCMD parameter constraints.

\label{subsec:ccmdmock} 
\begin{figure*}
\includegraphics[width=\textwidth]{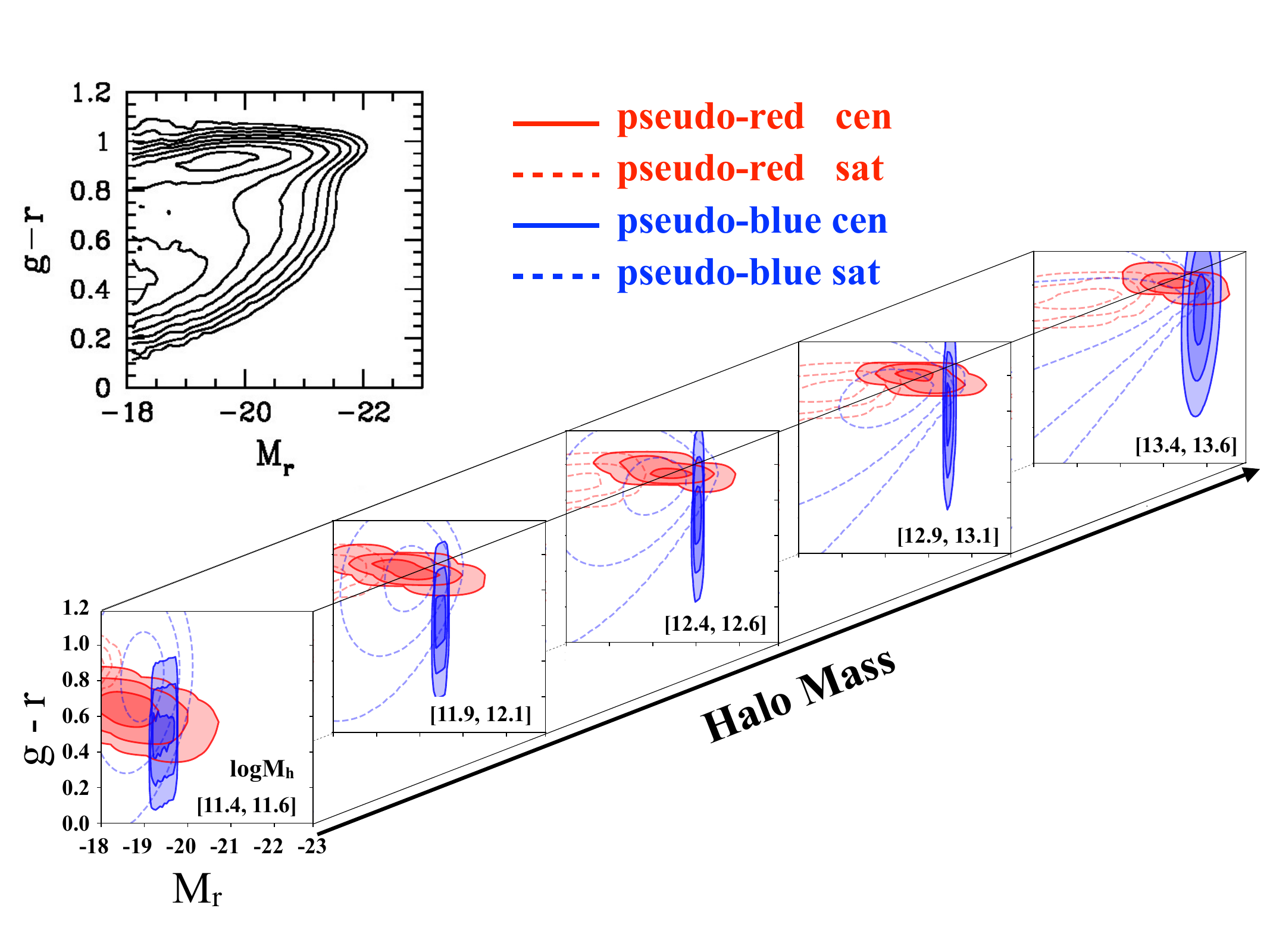}
\caption{
Illustration of the conditional colour--magnitude distribution (CCMD) formalism proposed in Xu18. The CCMD describes the distribution of
galaxy luminosity and colour as a function of halo mass.
The model consists of two populations motivated by galaxy colour bimodality, 
dubbed pseudo-blue and pseudo-red, respectively, 
with each further separated into central and satellite galaxies. 
A global parameterisation of these four colour-magnitude distributions
as a function of halo mass is developed, and the model parameters are determined by simultaneously ﬁtting the number densities 
and two-point auto-correlation functions of 79 galaxy samples deﬁned by ﬁne bins in the colour–magnitude diagram (CMD, top-left panel) of galaxies from the SDSS DR7. 
The model effectively deprojects the overall galaxy CMD along the halo mass direction (bottom cube), with the range of halo mass in terms of $\log[\Mh/(\hinvMsun)]$ labelled in each slice. Inside each slice, the contours describe the four CCMD components from the best-fitting model presented in Xu18 (see their fig.8). 
For each component, the contour levels are $\exp(-1/2)$,
$\exp(-4/2)$, and $\exp(-9/2)$ times the peak value for this component
(corresponding to the inclusion of 39, 86, and 99 per cent of galaxies in the
component for a 2D Gaussian distribution), and no overall normalisation is applied for the contour levels of different
components. 
}
\label{fig:cmd_ccmd}
\end{figure*}

\subsubsection{$N$-body simulation}
\label{subsubsec:nbody} 
The numerical $N$-body simulation CCMD mocks are based on is the
MultiDark MDPL2 simulation\footnote{\url{https://www.cosmosim.org/cms/simulations/mdpl2/}}. 
The MDPL2 simulation \citep{Klypin16} assumes a spatially flat cosmology
consistent with the constraints of
{\it Planck} \citep{PlanckCollaboration14,PlanckCollaboration16}, 
with $\Omega_{\rm m}=0.307$, $\Omega_{\rm b}=0.048$, $h=0.678$, $n_{\rm s}=0.96$, 
and $\sigma_8=0.823$. The simulation has a box size of 1 $h^{-1}{\rm Gpc}$ (comoving) 
on a side with $3840^3$ particles, with a mass resolution
of $1.51\times 10^9 \hinvMsun$. 

The redshift zero halo catalogue is obtained by the phase-space halo finder
{\tt ROCKSTAR} \citep{Behroozi13}. It is the same halo catalogue used in Xu18 to model SDSS galaxy clustering to constrain the CCMD model with an accurate and efficient simulation-based method \citep{Zheng16}.
Haloes in this catalogue are host haloes,
whose centre is not within the virial radius of a more massive halo.

\subsubsection{Populating galaxies}
\label{subsubsec:mock_generate}

We paint $N$-body haloes with galaxies with $r$-band absolute magnitude $M_r$ and $g - r$ colour according to the virial mass of host haloes based on the best-fit CCMD modelling results in Xu18. 

Xu18 introduced the CCMD model to describe the distribution of galaxy luminosity and colour as a function of halo mass, as illustrated in Fig.~\ref{fig:cmd_ccmd}. 
At a given halo mass, the galaxy distribution in the colour--magnitude diagram (CMD)
is assumed to consist of four components, two for centrals (the pseudo-blue 
central and the pseudo-red central) and two for satellites
(the pseudo-blue satellites and the pseudo-red satellites).
Each central component is described by a 2D Gaussian distribution in absolute magnitude--colour space, while the absolute magnitudes of the satellite components are described by a Schechter-like function and their colours by a Gaussian distribution.
Characteristic quantities\footnote{For example, for a 2D Gaussian distribution describing the colour-luminosity distribution of a central component, the characteristic quantities include the 
centre positions, standard deviations in colour and in luminosity, and correlation strength 
between colour and luminosity.} of these four components are parameterised as a function of halo mass. 
These parameters are determined by simultaneously ﬁtting the number densities 
and auto-correlation functions of galaxy samples defined in $\sim$80 fine 
luminosity–colour bins from the SDSS DR7.
Xu18 demonstrated that the CCMD model can accurately
recover the observed joint dependence of clustering on colour and luminosity, as well as
the observed global CMD.

Given the galaxy samples used to constrain the CCMD model, the CCMD mocks are designed to best represent galaxies with $-22 < M_r < -18$ and $0 < g - r < 1.2$. To make the mocks more comprehensive, we have extended the model to include the most luminous galaxies (i.e. $M_r < -22$) and those with extreme blue and red colours (i.e. $g -r < 0$ and $g -r > 1.2$). 
Since only a small fraction of galaxies are in these ranges, the extrapolation does not practically affect the accuracy of the mocks and any of our results. 
Central galaxies are assumed to reside at the potential
minimum of their host haloes and inherit their bulk velocities. 
For the satellite galaxies, we assign them the positions and velocities
of randomly chosen dark-matter particles.
Neither centrals nor satellites have assumed a velocity bias
\citep[e.g.][]{Guo2015a,Guo2015b}.
In each halo, the absolute magnitude $ M_r$ and $g -r$ colour of central galaxies are drawn from a 2D Gaussian distribution specified by its mass. While for satellites, we first calculate the mean number of satellites with $ M_r < -18$, based on the satellite CLF in haloes of this mass. The exact number of satellites for this halo is then drawn from a Poisson distribution with the mean. The absolute magnitude $ M_r$ is assigned according to the cumulative CLF of the satellite galaxies. The satellite $g -r$ colour is drawn from a Gaussian distribution following the CCMD parameterization.
It is possible that satellites are more luminous than their centrals. In the CCMD mock, we find that the fraction of haloes whose brightest galaxy is a satellite is a function of halo mass, 2\% for haloes with virial mass $\sim 10^{12} \hinvMsun$, 10\% for $\sim 10^{13} \hinvMsun$, and 12\% for $\sim 10^{14} \hinvMsun$, in agreement with the trend found in \citet{Skibba2011} and \citet{Lange2018}.
In conclusion, the CCMD mocks contain galaxies with realistic magnitudes and colours, all of which are more luminous than $M_r = -18$.

In addition to the mock catalogue based on the best-fitting CCMD parameters, we also construct 20 mocks from randomly selected CCMD models from the MCMC chain (Xu18) to address the effect of CCMD model uncertainties. The halo catalogue and the CCMD galaxy mocks are publicly available at
\url{https://www.astro.utah.edu/~zhengzheng/data.html}.

\subsubsection{Mock redshift surveys}
\label{subsubsec:octant}

We construct galaxy redshift surveys from the CCMD mocks starting by placing a virtual observer at a random position in the simulation box. 
To make a full-sky cone mock that reaches the maximum redshift $z \sim 0.2$, we periodically stack the cube mock to a big cube of 3 $h^{-1}{\rm Gpc}$ on a side.
We then compute the right ascension, declination, redshift, and apparent $r$-band magnitude $m_r$ for each mock galaxy, taking into account the peculiar velocity along the line of sight of the virtual observer and using the same cosmology parameters as the $N$-body simulation ($\Omega_{\rm m}$ and $h$) in conversion from comoving distance to redshift.
We exclude galaxies with $m_r > 17.6$ (the apparent magnitude
limit of NYU-VAGC {\it bright1}) and only keep galaxies within the redshift range $0.01 \leq z < 0.2$. 
We then carve out the SDSS footprint to mimic the edge effect.
We assign fibre collision status to mock galaxies following the method described in \citet{Guo2012}, which makes the fraction of collided galaxies match the observed one, see also \citet{Wang2022}. 
As with SDSS galaxies, we assign the redshift of the nearest neighbour in angular distance to each collided galaxy and calculate its absolute-magnitude $M_r$ and $g-r$ colour accordingly.
Given that only approximately $0.1\%$ galaxies have a spectral incompleteness less than 0.8 in our data, we discard those CCMD galaxies with low spectral incompleteness, i.e. $f_{\rm got} < 0.8$.
Finally, we apply the group finder to mock redshift surveys.

To make the most of the full-sky CCMD mocks and to consider the potential effect of sample variance, we carve 4 copies of the SDSS footprint from the full-sky CCMD cones and apply the same group finder to these quandrants.
We then calculate the mean and standard deviation of all relevant quantities in the CCMD groups from the quandrants.

\subsection{Group finder}
\label{subsec:gf} 

We choose the widely used $\Yang$ group finder to carry out the
group-finding process for all galaxy samples, including both CCMD mocks and SDSS DR7. 
The $\Yang$ group finder is a halo-based group finder and has been
tested and successfully applied to galaxy samples with spectroscopic
\citep{Yang05b, Yang07} and/or photometric redshift \citep{Yang2021}.
We shall refer to the group finder as the $\Yang$ group finder for short, unless otherwise specified.
In the following part, we give a brief description of how the
group-finding algorithm groups galaxies and assigns the group masses and memberships: 
\begin{enumerate}
   \item Start by assuming that each galaxy is a tentative group.\label{item:1}
   \item Measure the group luminosity of tentative groups. The group luminosity is the total luminosity of all member galaxies. \label{item:2}
   \item Assign each group a group mass $M_g$ by abundance-matching the cumulative group
   luminosity function and pre-chosen halo mass function at the sample redshift. 
    Record the mass-to-light ratios. When calculating the cumulative group luminosity function, weight groups by $1/V_{\rm BGG, max}$, the maximum volume where the brightest member galaxy can be observed.
    We use the \citet{Tinker08} halo mass function with the mass definition $M_{\rm 200b}$, where $M_{\rm 200b}$ defines the halo mass within the sphere whose mean density is 200 times that of the background universe. We use the same cosmology as in the MDPL2 simulation.\label{item:3}  
   \begin{enumerate}
     \item Recompute the group mass by the group luminosity and mass-to-light ratios from step \ref{item:3}. Then compute the halo radius from the halo mass within the sphere whose mean density is 180 times that of the background universe. 
     Assuming the virial theorem, compute the velocity dispersion of dark matter within a halo by halo mass and halo radius\footnote{We refer the interested reader for the detailed formulas in \citet{Yang05b,Yang07,Yang2021}.}. \label{item:3a}
     \item Determine the galaxy memberships. Assuming that dark matter haloes follow an NFW density profile \citep{NFW97} and that galaxies follow the same distribution as dark matter in phase space, the number density contrast of galaxies in the redshift space relative to the group centre (the positions of brightest member galaxy in the tentative groups) can be written as $P_M(R, \Delta z)=\frac{H_0}{c} \frac{\Sigma(R)}{\bar{\rho}} p(\Delta z)$, where $\Sigma(R)$ is the 2D projected density profile of the NFW profile at projected distance $R$ to the group centre.
     $\bar{\rho}$ is the average density of the universe, and $\Delta z=z-z_{\text {group}}$. $p(\Delta z)$ describes the redshift distribution of member galaxies by assuming that the velocity dispersion of satellites follows that of dark matter. 
     For each galaxy, loop over all groups to compute the distance pairs $(R, \Delta z)$ to all group centres. Assign the galaxy to the group(s) with $P_M(R, \Delta z) \ge B$, where $B$ is a pre-chosen parameter that tuned from galaxy mocks. 
     We adopt $B=10$ the same number as used in the widely used SDSS group catalogues. If a galaxy can be assigned to more than one group, assign it to the one with the highest $P_M(R, \Delta z)$ value. If all members of one group can be assigned to another group, merge these two groups into one group.
     \item With new determined memberships, we redetermine the group centre (the luminosity-weighted centre) and remeasure the group luminosity.
     \item Go back to step \ref{item:3a}. Iterate until there are no further changes to the group membership.
     \label{item:3d}
   \end{enumerate}
   \item With the memberships from step \ref{item:3d}, go back to step \ref{item:2} and iterate. Iterations stop when the mass-to-light ratios have converged, which typically takes three to four iterations.
 \end{enumerate}
Note that the algorithm presented above is adapted from \citet{Yang2021}, which
is slightly different from the classical one presented in \citet{Yang05b, Yang07}, and is optimised for large deep redshift surveys including photometric redshift data.
The changes are quite minor, mainly in steps~\ref{item:1} and ~\ref{item:2}. 
For our study, we do not need the optimal group finder, as long as the same one is used for both the CCMD mocks and the SDSS data. 
We have verified that there are no significant differences in the colour--magnitude--halo mass relation of \citet{Yang2021} and \citet{Yang05b, Yang07}.

\section{CCMD Groups VS. SDSS Groups}
\label{sec:CCMDgroups_SDSSgroups}

\begin{figure*}
\includegraphics[width=\textwidth]{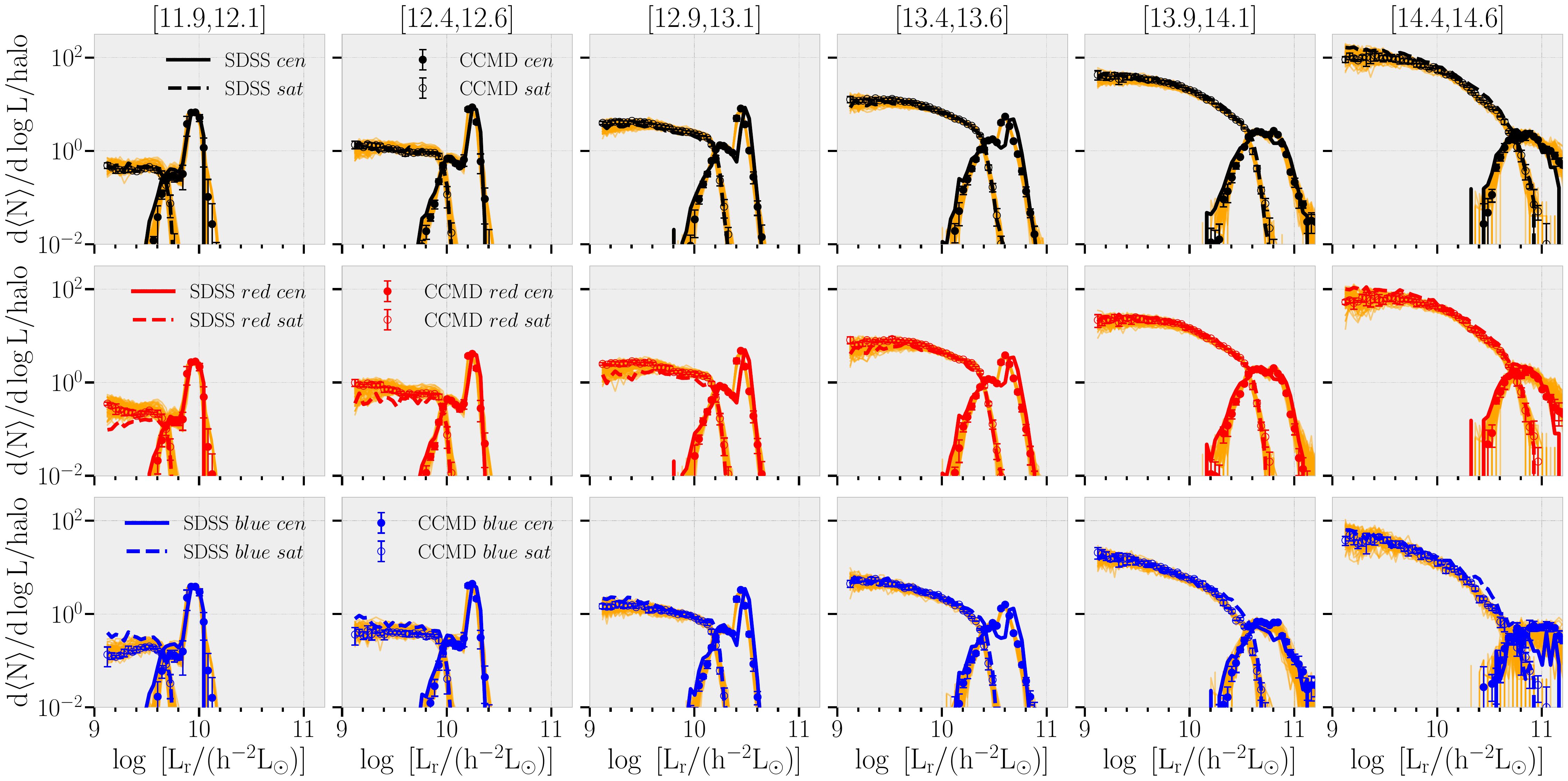}
\caption{
Comparisons of CLFs from the CCMD groups and the SDSS groups. The range of group mass in terms of $\logMg$ is labelled at the top of each column. The SDSS and CCMD groups are obtained by applying the $\Yang$ group finder to the SDSS DR7 data and the CCMD mock redshift surveys, respectively. 
The CLF (top row) is broken down into that of red galaxies (middle row) and that of blue galaxies (bottom row). 
The CLFs with errorbars from the CCMD groups come from the mock constructed using the best-fitting CCMD model, while the yellow curves are from the CCMD mocks constructed using randomly sampled CCMD parameters, representing CCMD model uncertainties. See Section~\ref{sec:CCMDgroups_SDSSgroups} for more details.
}
\label{fig:CLF_CCMD_groups_SDSS_groups}
\end{figure*}

With galaxy groups identified in the SDSS data and in the CCMD mocks using the same group finder, it becomes possible to make fair comparisons of galaxy properties in SDSS groups and CCMD groups. 

Groups serve as a proxy to dark matter haloes, but they are far from perfect tracers of haloes.
There are at least three
types of failure that a group finder can experience (following the naming convention in Campbell15): membership allocation errors, central/satellite designation errors, and halo mass estimation errors.
In most cases, these errors are coupled to each other. 
When using a group-finding algorithm to group galaxies, 
two partition errors can occur: fracturing and fusing. 
Fracturing occurs when galaxies that belong to the same halo
are incorrectly identified as two or more groups.
Fusing is when galaxies from multiple haloes are mistakenly grouped into one. 
These two issues often occur simultaneously, for example, when galaxies from two distinct but nearby haloes are divided into two groups, with some of the galaxies from one halo being placed in one group (``fracturing'' for this halo) and the rest being incorrectly absorbed into the other group (``fusing'' for the other halo). See Fig.3 in Campbell15 for an illustration.
Even if all member galaxies are correctly identified, errors in the designation of central and satellite galaxies can occur --- the algorithm usually chooses the brightest member as the central galaxy, while there is a considerable fraction of haloes whose central galaxies are not the brightest \citep[e.g.][]{Skibba2011, Lange2018}. 
Furthermore, even if memberships and centrals are correctly identified, the halo mass estimation can be inaccurate due to the scatter and dependence on secondary properties such as galaxy colours \citep[e.g.][]{Mandelbaum2016, Xu2018}.

Given the nature of a group finder, groups would not provide the exact galaxy colour-luminosity-halo mass relation, as inferred from Xu18. However, applying the same group finder to the SDSS data and to the CCMD mocks results in the same type of group-finding and group mass-assignment systematics in the SDSS groups and CCMD groups, making fair comparisons.

We can treat galaxy groups from a group-finding algorithm as a way to define the environment, and the comparisons are for environment-dependent galaxy properties. As long as the environment is defined in the same way in the data and in the mocks, the comparisons are meaningful. In this sense, the specific group-finding algorithm does not matter for our study here. In Appendix~\ref{sec:Tinker2020}, we perform a test by applying a different group finder (i.e. the {\tt Tinker2020} group finder; \citealt{Tinker2020,Tinker2021}) to both CCMD mocks and SDSS data. We find that our comparison results are consistent with those from the default group finder that we adopt.

We make comparisons of CLFs, conditional colour functions (CCFs), and CCMDs from the SDSS and CCMD groups, as a function of group mass. When counting galaxies and groups from either the SDSS data or the CCMD mock redshift survey (corresponding to an octant of the CCMD mock; see Section~\ref{subsubsec:octant}), we weigh galaxies by $1/V_{\rm max}$ and groups by $1/V_{\rm cen, max}$, where $V_{\rm cen, max}$ is the $V_{\rm max}$ of the central galaxies of the groups. For CLFs and CCFs from the CCMD groups, we estimate the uncertainties from the dispersions of the corresponding quantities among the four quadrants of the full-sky CCMD spheres (each mimics SDSS footprints and fibre collisions; see Section~\ref{subsubsec:octant} for details). To access the uncertainties of the CCMD model, we also measure CLFs and CCFs from 20 cubic mocks populated with CCMD parameters randomly sampled from the CCMD modelling posteriors from Xu18. We apply the same procedures to these mocks as done in Section~\ref{subsubsec:octant} to obtain realistic SDSS galaxy mocks. Therefore, in the relevant plots, the corresponding CLFs and CCFs measured from these 80 mock surveys are plotted as orange curves.

\subsection{The conditional luminosity function (CLF)}

\begin{figure*}
\includegraphics[width=\textwidth]{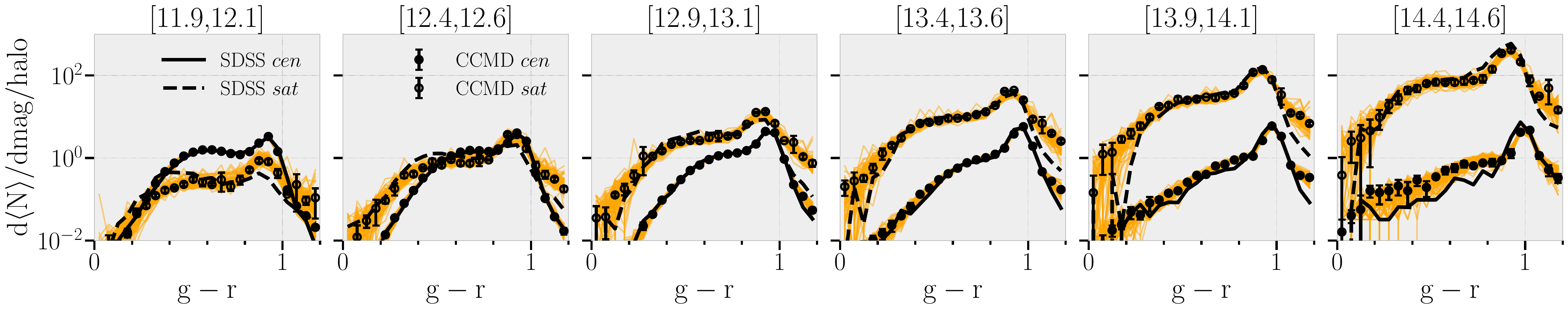}
\caption{
Similar to Fig.~\ref{fig:CLF_CCMD_groups_SDSS_groups}, but for comparisons of CCFs between the SDSS and CCMD groups. 
}
\label{fig:CCF_CCMD_groups_SDSS_groups}
\end{figure*}

\begin{figure*}
\includegraphics[width=\textwidth]{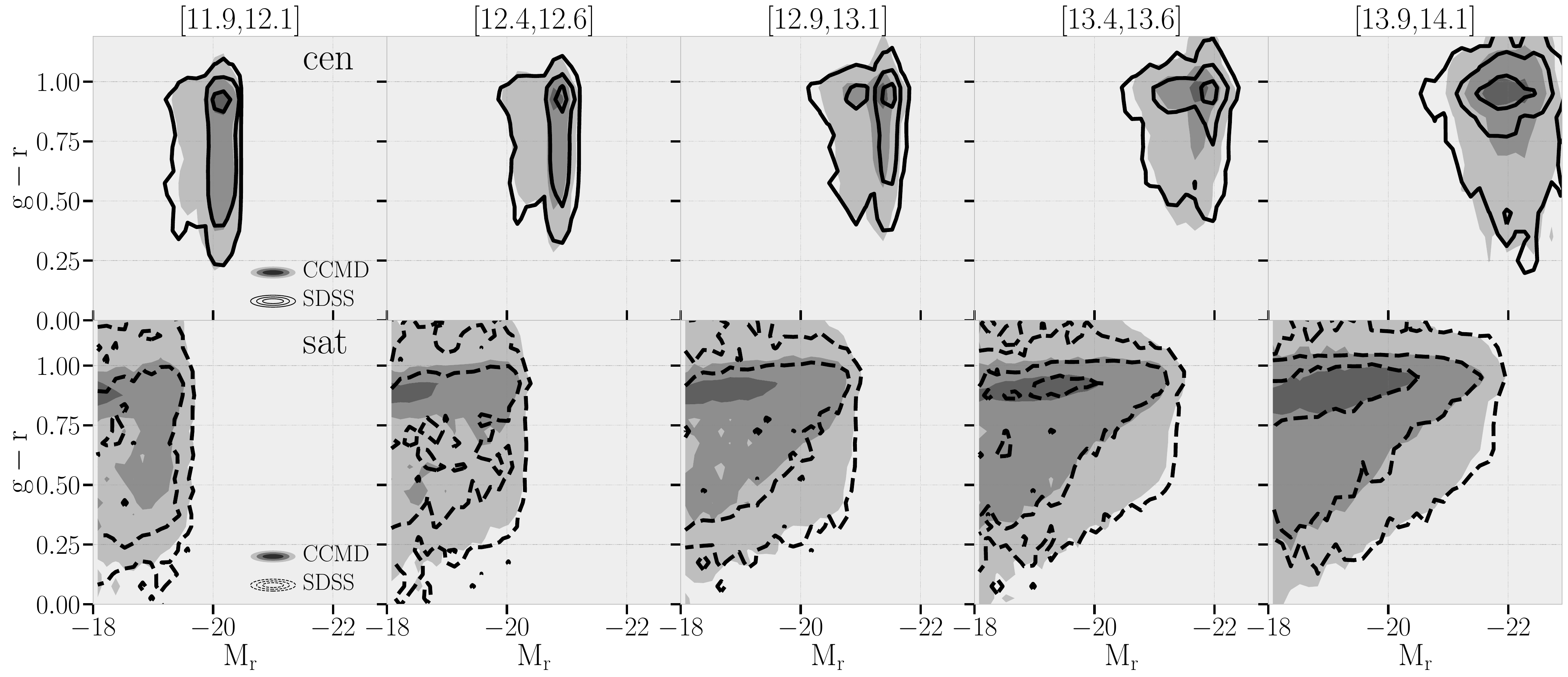}
\caption{
Comparisons of CCMDs between the CCMD groups and SDSS groups, separated into central CCMDs (top) and satellite CCMDs (bottom). 
In each panel, the contour levels are $\exp(-1/2)$, $\exp(-4/2)$, and $\exp(-9/2)$ times the maximum value in the CCMD of the SDSS groups. Note that the CCMD groups for the comparisons are obtained from the CCMD mock constructed with the best-fitting CCMD parameters, and the uncertainties in the CCMD model are not reflected here.
}
\label{fig:CCMD_CCMD_groups_SDSS_groups}
\end{figure*}

We first compare the CLFs, the galaxy luminosity distribution as a function of group mass, between the CCMD groups and the SDSS groups in Fig.~\ref{fig:CLF_CCMD_groups_SDSS_groups}. 

In the top panels, the central galaxy CLFs from the SDSS groups (solid curves) and from the CCMD groups (filled circles) appear to be remarkably similar, given that the amplitude of the CLFs can vary by about 3 orders of magnitude at fixed group mass. Except for the highest mass bin, the central CLFs from both SDSS and CCMD groups show two peaks, with the lower-luminosity peak being broader. The higher-luminosity peak from the CCMD groups shifts toward slightly higher luminosity than that from the SDSS groups. 

While it is tempting to attribute the double peaks as an indication of the two central components in the CCMD (Xu18), our further investigations suggest an origin more related to the effect of systematics in group mass assignment. As pointed out by \citet{Yang2005c, Yang08}, a sharp feature in the central CLF can result from the halo mass assignment based on the ranking of group luminosity, particularly for low-mass haloes, which are mainly composed of central galaxies. This issue can be addressed in deep surveys \citep[e.g.,][]{Wang2023} or by using the total group stellar mass as a halo mass indicator instead. A more detailed discussion can be found in the Appendix~\ref{sec:clf_double_peak}. 
We emphasise here that the focus of this work is not the exact shape of the CLFs, but the comparison of group-based statistics between CCMD groups and SDSS groups. 
The exact shape of the CLFs (for example, the double-peak feature) is sensitive to the group finders (c.f. Appendix~\ref{sec:Tinker2020}), the evaluation volume, and the bin widths for mass and luminosity (c.f. Appendix~\ref{sec:clf_double_peak}).

If we further divide galaxies into red and blue populations using a magnitude-dependent cut, $(g-r)_{\rm cut} = 0.21-0.03M_r$ \citep{Zehavi11}, the central CLFs of the two populations (middle and bottom panels) between SDSS and CCMD groups also show a level of agreement similar to the total population.

In the top panels, the satellite CLFs from SDSS groups (dashed curves) and CCMD groups (open circles) also show remarkable agreements, in both the slope and amplitude toward the low luminosity end and the cutoff toward the high luminosity end. 
However, when satellites are divided into blue and red populations, the CLF for blue satellites from CCMD groups with the lowest group mass $\logMg \sim 12$ (in the leftmost panel) is underestimated by about 50 per cent, and that for red satellites is overestimated by about 50 per cent. CCMD model uncertainties indicated by the yellow curves, estimated using mocks constructed by randomly selected 20 CCMD models in the MCMC chain, can help explain the discrepancy between satellite CLFs in the low-mass SDSS and CCMD groups. 

As a whole, except for the red and blue satellite CLFs in low-mass groups, the central and satellite CLFs from SDSS groups and CCMD groups show excellent agreement, implying that the CCMD model in Xu18 provides a reasonable description of the galaxy luminosity-halo mass relation.

\subsection{The conditional colour function (CCF)}

Here, we define the conditional colour function (CCF) as the colour distribution of member galaxies at a fixed group mass. CCF comparisons are shown in Fig.~\ref{fig:CCF_CCMD_groups_SDSS_groups}.

For central galaxies, the CCFs from the SDSS groups (solid curve) agree with those from the CCMD groups (filled circles). At each group mass, the CCF looks like a superposition of two Gaussian-like distributions, a diffuse distribution centred at a blue colour and a narrow distribution at a red colour. Such features are a reflection of the two central components (pseudo-blue and pseudo-red) in the CCMD model (Xu18). In groups of the three intermediate mass range,
$\logMg \sim 13.0$, 13.5, and 14.0, the CCFs from the CCMD groups have a more pronounced red tail, overpredicting central galaxies with colour $g-r>1.1$. As there are only a small number of extremely red galaxies in the SDSS data, it is not surprising that the CCMD model does not accurately capture the red tail of the distribution. Furthermore, the red tail in the CCMD model is largely from the tail of the pseudo-blue central component, which is described as a Gaussian distribution. The result here suggests that we may need to consider a more sophisticated functional form to have a more accurate description of the colour distribution of this component for an improved CCMD model.

Similar to the central CCFS, the satellite CCFs from both SDSS and CCMD groups (dashed curves and open circles) display a bimodal distribution. As in the CLF comparison, the CCMD model underpredicts the number of blue satellites in low-mass groups. Like the central CCFs, the CCMD model overpredicts the red tail of the satellite CCFs in high-mass groups, suggesting the need of an improved functional form of the pseudo-blue satellite colour distribution.

The overall trend of the CCF from the SDSS groups, separated into those for centrals and satellites, is well reproduced by the CCMD groups. The mismatch of satellite CCFs in low-mass groups and the overprediction at the red tail in both central and satellite CCFs from the CCMD groups suggest a possible direction to improve the CCMD model.

\subsection{The conditional colour-magnitude distribution (CCMD)}

The CLF and the CCF are two projections of the CCMD. The consistency between CLFs and CCFs from the SDSS and CCMD groups indicates good agreements between CCMDs from the data and the mock. We turn to direct comparisons of group CCMDs.

In Fig.~\ref{fig:CCMD_CCMD_groups_SDSS_groups}, the CCMDs from the SDSS groups (contours) and the CCMD groups (grey scales) are compared in groups with mass ranging from $\logMg\sim$ 12 to 14, separated into those of central and satellites.

For central galaxies, the results in Xu18 show that the CCMD at fixed halo mass has two components, one with narrow luminosity distribution but spread in colour (the pseudo-blue component) and the other with narrow colour distribution but spread in luminosity (the pseudo-red component). The central CCMD at fixed group mass from the SDSS groups and CCMD groups looks like a superposition of two components. As with the CLF, the component with a narrow luminosity distribution has a large contribution from those groups with a single member galaxy identified (see Appendix A). We performed tests to see the effect caused by such groups by removing them in computing the CCMD. We found that the CCMD still looks like a superposition of two components, while the redder component does not have a luminosity distribution as spread as the halo CCMD (Xu18). Therefore, it seems that the group CCMDs keep the two-component features in the halo CCMDs but with a substantial smearing effect, resulting from group-finding algorithms (such as member determination, central/satellite separation, and group mass assignment). 

The overall central galaxy CCMDs from the SDSS groups and the CCMD groups display a general agreement. In groups of mass $\logMg\sim $ 12.5, 13.0, and 13.5, the CCMD groups tend to have central galaxies about 0.2 mag fainter than the SDSS groups, specifically at the luminous end.

For the satellite CCMD, the overall patterns for the SDSS groups and CCMD groups are similar, showing a relatively tight red sequence and a diffuse blue component. In detail, at low group mass, $\logMg\lesssim 13$, there are slightly fewer satellites in the CCMD groups. At high group mass, the agreement is remarkably good, except that the CCMD groups have more satellites at the red tail. The comparisons are in line with those seen in the CCF comparisons (Fig.~\ref{fig:CCF_CCMD_groups_SDSS_groups}).

\medskip

The results of the CLF, CCF, and CCMD comparisons between the CCMD groups and the SDSS groups are encouraging. The generic agreements suggest that the CCMD model in Xu18 provides a good description between galaxy colour-luminosity distribution and halo mass. 

\begin{figure*}
\includegraphics[width=\textwidth]{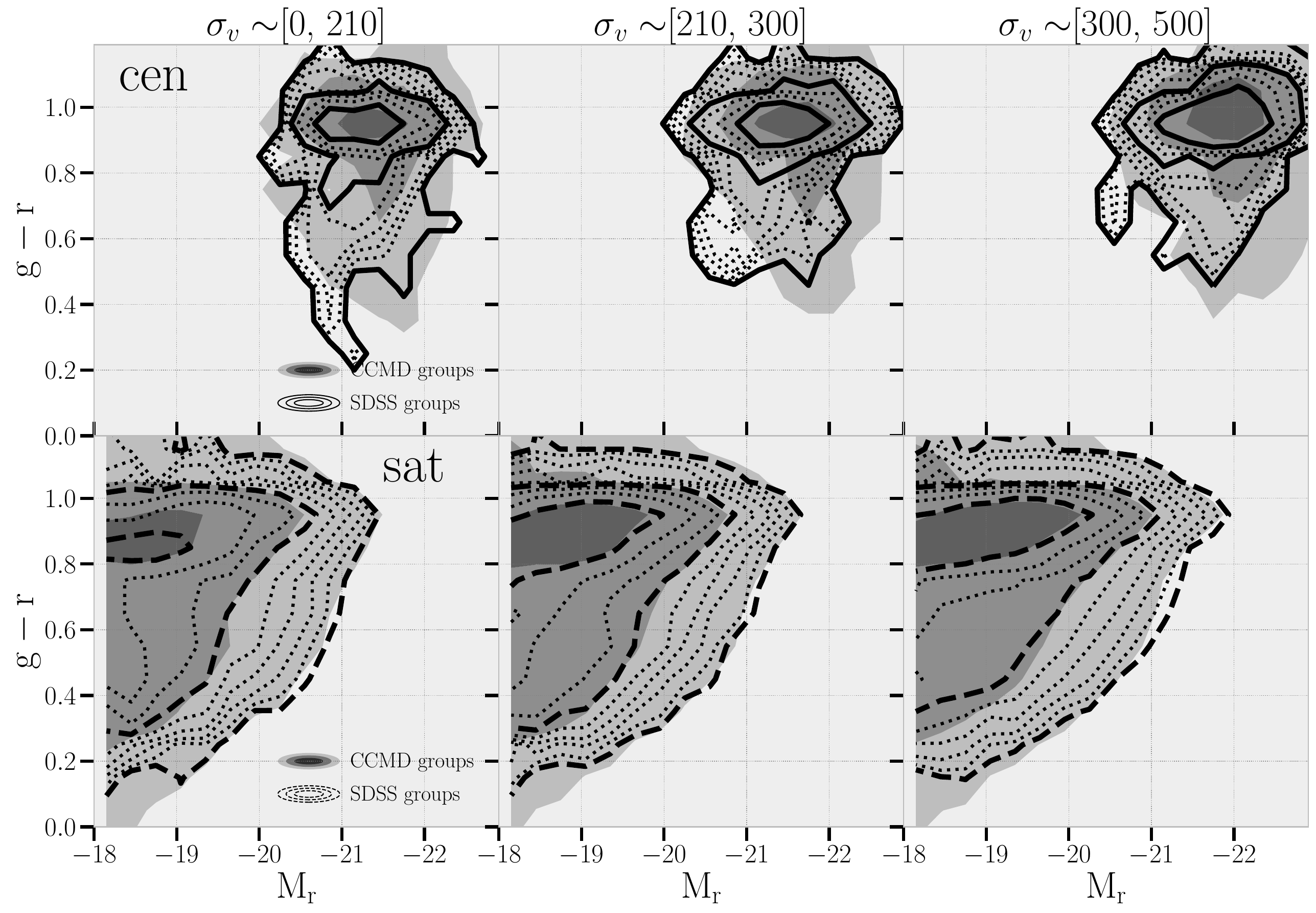}
\caption{
Comparisons of CCMDs between the CCMD groups and the SDSS groups, as a function of line-of-sight velocity dispersion ($\sigma_v$) of satellite galaxies. The range of $\sigma_v$ in ${\rm km\, s^{-1}}$ is labelled at the top of each column.
In each panel, for SDSS groups and CCMD groups, the thick contour and greyscale levels are $\exp(-1/2)$, $\exp(-4/2)$, and $\exp(-9/2)$ times the maximum value of the distribution from SDSS groups. To further highlight the details of the CCMD from the SDSS groups, additional contour levels (thin dotted curves) are plotted.
}
\label{fig:ccmd_vdisp}
\end{figure*}

\section{Additional CCMD Comparisons}
\label{sec:direct_ccmd}


In this section, we attempt to carry out further investigations of the CCMD features from observation. Galaxy groups identified by the group finder have systematic errors, such as group mass assignment and membership allocation. Compared to the CCMD at fixed halo mass, the CCMD at fixed group mass has certain features smeared out. To reduce the systematics related to group mass assignments, in Section~\ref{subsec:sat_vdisp}, we make comparisons by dividing the groups by their dynamical group mass based on the velocity dispersion of satellite galaxies. 
In addition to group mass assignment systematics, we also attempt to reduce central/satellite designation systematics. 
In Section~\ref{subsec:cen_cat}, we investigate the colour and luminosity distribution of
``true'' central galaxies extracted from two external central galaxy catalogues. The mean host halo mass of these central galaxies has been calibrated using weak-lensing measurements.

\begin{figure*}
\includegraphics[width=\textwidth]{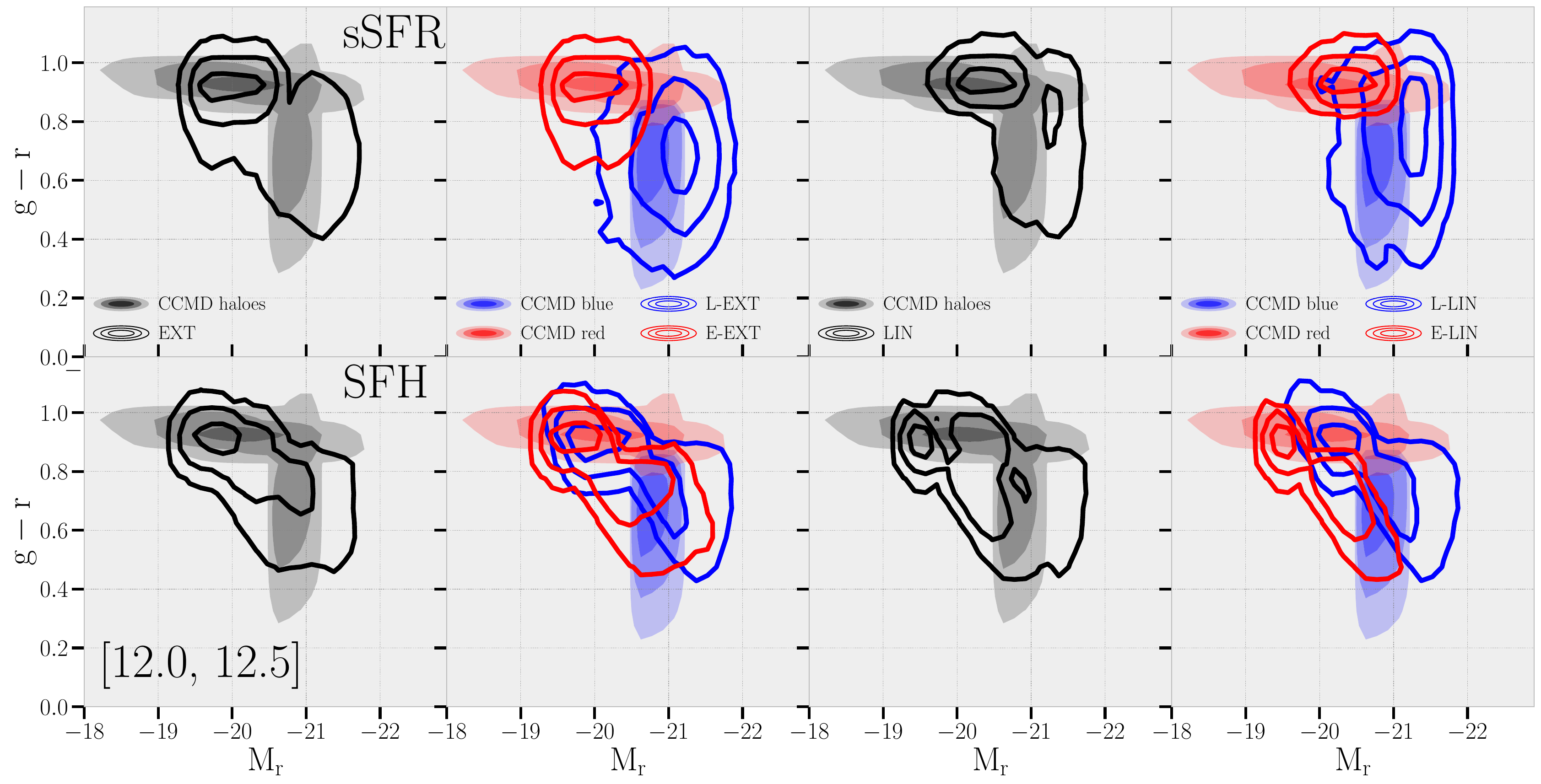}
\caption{
Comparisons of CCMDs of central galaxies in CCMD haloes to those from two catalogues of SDSS central galaxies, EXT \citep{McCarthy2022} and LIN \citep{Lin16}. Galaxies from the central galaxy catalogues are divided into early-forming (E-EXT and E-LIN) and late-forming (L-EXT and L-LIN) subsamples, based on star formation rate (sSFR; top panels) and star formation history (SFH; bottom panels). The mean host halo masses of those central galaxies are estimated to be in the range of $\sim 12 < \logMh < 12.5$ from weak lensing measurements, which is also the mass range of the haloes chosen to show the central CCMD from the CCMD mock. In the panels in the second and fourth columns, the central galaxies are split into blue and red populations.
Note that the contour levels in this figure are $\exp(-1/2)$, $\exp(-4/2)$, and $\exp(-9/2)$ times the maximum value of the distribution of each component, i.e. no overall normalisation is applied for the contour levels of different components.
Note that the mass used by LIN and EXT is $M_{\rm 200c}$, which defines the halo mass within the sphere whose mean density is 200 times the critical density of the universe. For fair compassion, we have converted the viral mass given by the CCMD halo catalogue to $M_{\rm 200c}$ by assuming an NFW profile and applying the mass-concentration relation from \citet{Diemer2019}.}
\label{fig:CCMD_cen_cat}
\end{figure*}

\subsection{CCMD conditioned on satellite velocity dispersion}
\label{subsec:sat_vdisp}
For a group with sufficient number of galaxies, the velocity dispersion $\sigma_v$ of satellites can be estimated. It is a measure of the dynamical mass of the group, which can be a good proxy for the underlying dark matter halo mass (approximately $\Mh\propto \sigma_v^3$). This motivates us to examine the CCMD as a function of satellite velocity dispersion. 

We estimate the line-of-sight velocity dispersion from satellite galaxies for each group that contains more than 10 members using the gapper estimator\footnote{We have verified that using the direct estimator from the variances of the satellite velocities leads to almost the same results.} \citep{Beers1990,Yang05b,Yang07}.
The method first sort the recession velocity $v_i \equiv cz_i$ of $N$ group members (ascending order), where $c$ is the speed of light. And the gaps are defined as,
\begin{equation}
g_i = v_{i+1} -v_i, i = 1,2,...,N-1.
\end{equation}
Then the rest-frame velocity dispersion is estimated by
\begin{equation}
\sigma_{\rm gap}=\frac{\sqrt{\pi}}{\left(1+z_{\rm group}\right) N(N-1)} \sum_{i=1}^{N-1} w_i g_i,
\label{eq:vdisp}
\end{equation}
where the weight $w_i$ is defined as $w_i = i(N-i)$. 
Since the central galaxy is assumed to be at rest with respect to its host group, the final velocity
dispersion should be corrected as
\begin{equation}
\sigma_v = \sqrt{\frac{N}{N-1}} \sigma_{\rm gap}.
\end{equation}
In Eq.~\ref{eq:vdisp}, we take $z_{\rm group}$ as the mean redshift of all group satellites, 
and we have verified that the CCMD results remain unchanged if we replace it with the redshift of the brightest member galaxy or the luminosity-weighted redshift of all group members.

The CCMD as a function of the line-of-sight velocity dispersion is shown in Fig.~\ref{fig:ccmd_vdisp}. To have sufficient statistics, we divide the groups into three broad bins, with satellite velocity dispersion $\sigma_v\sim$ [0, 210], [210, 300], and [300, 500] ${\rm km\, s^{-1} }$, respectively. If we assume satellites follow the velocity of dark matter particles (i.e., with velocity bias neglected), the three ranges approximately correspond to halos of mass $\logMh\sim$ below 13.0, between 13 and 13.5, and between 13.5 and 14.1, respectively.

For the central CCMDs, those from the CCMD groups appear to have a slight shift toward higher luminosity, in comparison with those from the SDSS groups, which could imply the existence of a satellite velocity bias \citep[e.g.][]{Guo2015b}. In both cases, the central CCMD shows some hints of two approximately orthogonal components, with one component narrow in colour but spread in luminosity (indicated by the innermost contour) while the other spread in colour but narrow in luminosity (indicated by the outermost contour), which is consistent with the CCMD model. There is another noticeable difference between the CCMDs from the SDSS and CCMD groups --  the component with a narrow luminosity distribution is more prominent in the CCMD groups. Neglecting a velocity bias in central galaxies may contribute to such a difference, as the central velocity bias can affect the designation of central and satellite galaxies. 
To see the impact of velocity bias on group-based statistics, we generate a CCMD mock with velocity bias built-in and apply the observational effects as listed in Section~\ref{subsubsec:octant}. We adopt a velocity bias model from \citet{Guo2015b}, where central galaxies are not resting at the halo centres but have a velocity dispersion that is 25\% of that of dark matter, and satellite galaxies move slightly slower, with a velocity dispersion within the halo that is approximately 85\% of dark matter. This magnitude of velocity bias is driven by the redshift space 2PCFs measured from SDSS DR7. We then apply the same group finder to this velocity bias mock and find that it has almost negligible impact on the results presented here and in CLF/CCF/CCMD (see, e.g., Figs.~\ref{fig:CLF_CCMD_groups_SDSS_groups},~\ref{fig:CCF_CCMD_groups_SDSS_groups}, ~\ref{fig:CCMD_CCMD_groups_SDSS_groups}). 
Further investigation is needed to understand why the difference here is more significant than that seen in the comparison based on group mass (Fig.~\ref{fig:CCMD_CCMD_groups_SDSS_groups}).

Satellite CCMDs from the CCMD and SDSS groups show remarkable agreement. Note that the levels of inner contour and outer contour in each panel differ by a factor of about 50. As with the group-mass-based CCMDs, the CCMD groups have a more extended tail of red galaxies.

With groups divided by satellite velocity dispersion, the CCMD and SDSS groups show broad agreements in central CCMDs and good match in satellate CCMDs. The differences in the fine features of central CCMDs are worth a more detailed study. 

\subsection{CCMD based on catalogues of central galaxies}
\label{subsec:cen_cat}

The central/satellite designation systematics from any group finder can smear out features in the CCMDs. There are efforts to reduce such systematics by applying certain criteria to select central galaxies with identified galaxy groups \citep[e.g.][]{Lin16}. Here we attempt to make a comparison between the CCMDs from such central galaxies and the CCMD model in Xu18.

Xu18 found that at fixed halo mass the CCMD of central galaxies is composed of two orthogonal components: a red component that is narrow in colour and extended in magnitude, and a blue component that is narrow in magnitude and extended in colour. \citet{Xu2022} used semi-analytical models of galaxy formation and hydrodynamic simulations to investigate the CCMD of central galaxies and found evidence of these two components. We aim to look for this feature directly in the data. Ideally, we need a complete and pure catalogue of central galaxies, where all centrals are identified, and no satellite interlopers are present. Additionally, instead of the assigned group mass, it would be good to have the host halo mass.

\citet{Lin16} created two catalogues of central galaxies with similar halo masses
but different star-formation activities. They first selected central galaxies from the $\Yang$ group catalogue \citep{Yang07} and then removed satellites that could have
been misidentified as centrals. They divided central galaxies into early and late samples and applied a few further selection criteria with the intention that the two types of galaxies reside in haloes of similar mass. The mean halo masses of these two types of central galaxy samples are measured through weak lensing and are found to be in the range of $\logMh \sim$ 12.0 to 12.5. We name these catalogues LIN hereafter.
\citet{McCarthy2022} extend the LIN catalogues by increasing the halo mass range, and these extended catalogues were named EXT. The division of the early and late centrals in both the LIN and EXT catalogues is based on two criteria: star formation history (SFH) and specific star formation rate (sSFR).

In Fig.~\ref{fig:CCMD_cen_cat}, we compare the CCMD of central galaxies in the EXT (left two columns) and LIN catalogues (right two columns) with the CCMD of central galaxies in the CCMD mocks in haloes with a mass range of $12 < \logMh < 12.5$. The sSFR-based central galaxy samples show two orthogonal components (top row). 
When divided into late and early subsamples, the early subsamples in both catalogues are narrow in colour and extended in luminosity, while the late ones are narrow in luminosity and extended in colour. The broad features are consistent with the expectations from the CCMD model.
 
In the SFH catalogues (bottom row), two orthogonal components are still visible, though much less prominent than in the sSFR catalogues. Interestingly, the SFH-based early (late) subsample displays a blue (red) component, suggesting that colour is more closely associated with recent star formation (largely characterised by sSFR) than with the overall history of star formation.

In detail, the central galaxy CCMD from either the LIN and EXT catalogues with early/late galaxy division from either SFH or sSFR does not match the predicted one from the CCMD model (grey scale). This is not surprising, as there are systematics in defining the central galaxy samples, and the comparison is not an apples-to-apples comparison. In constructing central galaxy samples, it is still possible that the satellite decontamination step may have removed some true central galaxies and not all satellites were eliminated. While the {\it mean} halo masses of early and late galaxies are measured from weak lensing, they may have different distributions, which makes it less clear which halo mass range in the CCMD mock to use for the comparison. Furthermore, the CCMD model only has colour and luminosity information, and comparisons with sSFR- and SFH-based samples are not straightforward. With all of the caveats in mind, the two-component CCMD feature shown in the central catalogues lends support to the CCMD model.

\section{Discussions and Conclusions}
\label{sec:summary}

The CMD of galaxies exhibits a bimodal distribution that holds significant information about the formation and evolution of galaxies. To explore this distribution from the perspective of dark matter haloes, Xu18 developed the CCMD model. This model describes the colour and luminosity distribution of galaxies as a function of halo mass. It is composed of two colour populations, each of which is further divided into central and satellite galaxies. The model parameters were determined by simultaneously fitting the abundances and two-point auto-correlation functions of $\sim$80 galaxy samples defined by fine bins in the CMD from SDSS DR7. The CCMD extends the CLF framework by adding the colour dimension, providing a novel approach to probe the relationship between galaxies and dark matter haloes.

The results of the CCMD modelling can be compared with the galaxy-halo connections inferred from other methods. In particular, colour and luminosity distributions of galaxies in galaxy groups are closely related to CCMD, as groups are thought to be a good proxy for haloes, making them natural targets for comparing CCMD modelling results with those from galaxy groups. 
However, systematics such as incorrect membership assignment and inaccurate group mass assignment within a group-finding algorithm prevent a meaningful comparison, as shown in Xu18. To address this, we construct a mock galaxy catalogue by populating haloes in $N$-body simulations in accordance with the best-fitting CCMD model and then apply the same group finder to both the mock catalogue and the observational data. This ensures that both the CCMD groups (identified by the group finder in the CCMD mocks) and the SDSS groups (identified by the group finder in SDSS DR7) are subject to the same systematics. We also construct CCMD mock catalogues from a set of models randomly drawn from the CCMD posterior, and the corresponding CCMD groups from them are used to assess the model uncertainties in the comparison.

The comparison between CCMD groups and SDSS groups yields several key findings.
\begin{itemize}

    \item[(1)] The galaxy groups from the CCMD mocks are similar to those from the SDSS DR7 data in terms of CLF, CCF, and CCMD for central and satellite galaxies, regardless of the group finder used (cf. Figs.~\ref{fig:CLF_CCMD_groups_SDSS_groups}, ~\ref{fig:CCF_CCMD_groups_SDSS_groups}, ~\ref{fig:CCMD_CCMD_groups_SDSS_groups} and Figs.~\ref{fig:TinkerGF_CLF_CCMD_groups_SDSS_groups}, ~\ref{fig:TinkerGF_CCF_CCMD_groups_SDSS_groups}, ~\ref{fig:TinkerGF_CCMD_CCMD_groups_SDSS_groups}). 
    The agreements provide support for the validity of the CCMD modelling results. For groups with mass $\logMg < 12.5$, the CCMD mocks underestimate (overestimate) the number of faint blue (red) satellites. This can result from the limited constraining power for the CCMD in low-mass haloes, caused by the small survey volume for those galaxies.
    
    \item[(2)] When groups are binned according to the line-of-sight velocity dispersion of satellite galaxies, a proxy to halo mass, the central galaxy CCMD from the CCMD groups shows a slight shift with respect to that from the SDSS groups. The match in detailed distributions of colour and luminosity between the CCMD and SDSS groups is less impressive than that based on group mass, which we have ruled out the possibility of galaxy velocity bias (not included in the CCMD mocks).
    The satellite CCMDs, on the other hand, show good agreement between the CCMD and SDSS groups (cf. Fig.~\ref{fig:ccmd_vdisp}).

    \item[(3)] We also make use of catalogues of central galaxies selected by applying additional procedures to a galaxy group catalogue to remove satellite contamination, with the mean halo mass measured through weak lensing. The colour and luminosity distribution of central galaxies shows two distinct components that are orthogonal to each other, a red component that is narrow in colour but wide in luminosity, and a blue component that is restricted in luminosity but broad in colour. Such a two-component distribution agrees with the predictions of the CCMD model (cf. Fig.~\ref{fig:CCMD_cen_cat}).

    \item[(4)] It has been demonstrated that the group finder, which determines the mass of a group based on the total luminosity of its member galaxies, would generate a false double-peak feature in the CLF of central galaxies, regardless of the input CLF shape (cf. Fig.~\ref{fig:CLF_pbcen_removed}). The feature can be hidden when the CLF is evaluated using galaxies in a limited volume and/or with a wide group mass and luminosity bin (cf. Figs.~\ref{fig:CLF_double_peak},~\ref{fig:CLF_Meng2022}).
    
\end{itemize}

As a follow-up study of Xu18, this work aims to have an apples-to-apples comparison between the galaxy colour--magnitude--halo mass relations inferred from Xu18 and from the group catalogues. The results suggest that the CCMD mocks provide a good representation of the galaxy distribution in the universe and the galaxy colour and luminosity distribution of galaxies in haloes. 

The differences revealed in the comparison need further investigation, including the slight shift in the central CCMD, lower amplitude in satellite CLFs in low-mass groups from the CCMD mock, extended red tail of galaxies in the CCMD mock, and fine details in the central CCMD based on satellite velocity dispersion. These can be related to cosmology (e.g. redshift of simulations used in constructing mocks), insufficiency in the CCMD parameterisation, limited constraining power from galaxy surveys for low-luminosity galaxies.

Meanwhile, realistic mock catalogues, such as those constructed in this work, allow for an array of investigations. For example, they can be used to test the systematics of group-finding algorithms and help improve the performance of group finders. As the mocks use halo mass as the sole variable to determine the galaxy content in haloes, no assembly bias effect \citep[e.g.][]{Gao2005} is built in. Comparing certain statistics in the mock and in the data can help constrain galaxy assembly bias. For instance, they can be applied to test the galactic conformity \citep[e.g.][]{Weinmann2006,Campbell15,Calderon2018, Tinker2018}, separating the true effect (if there is any) from that artificially induced by the systematics of group finder. We reserve such investigations for future work. CCMD modelling with data from large galaxy surveys, such as the Dark Energy Spectroscopic Instrument (DESI; \citealt{DESI}) survey, will lead to more investigations and applications and allow stringent comparisons to help understanding galaxy formation and evolution.

\section*{Acknowledgements}
The authors thank Kevin S. McCarthy and Yen-Ting Lin for supplying their collections of central galaxies. We thank Kuan Wang, Houjun Mo and Jiaxin Han for helpful discussions. HX expresses appreciation to the Department of Astronomy at Shanghai Jiao Tong University for their hospitality. The support and resources from the Center for High Performance Computing at the University of Utah are gratefully acknowledged. This work is supported by the National SKA Program of China (grant No. 2020SKA0110100), National Science Foundation of China (Nos. 11833005, 11890692, 11922305), 111 project No. B20019, and Shanghai Natural Science Foundation, grant No.19ZR1466800. We acknowledge science research grants from the China Manned Space Project with No.CMS-CSST-2021-A02. HG acknowledges the support from the CAS Project for Young Scientists in Basic Research (No. YSBR-092) and GHfund C(202407031909). We acknowledge the use of the Gravity Supercomputer at the Department of Astronomy, Shanghai Jiao Tong University.

The CosmoSim database used in this paper is a service by the Leibniz-Institute for Astrophysics Potsdam (AIP). The MultiDark database was developed in cooperation with the Spanish MultiDark Consolider Project CSD2009-00064. The authors gratefully acknowledge the Gauss Centre for Supercomputing e.V. (www.gauss-centre.eu) and the Partnership for Advanced Supercomputing in Europe (PRACE, http://www.prace-ri.eu) for funding the MultiDark simulation project by providing computing time on the GCS Supercomputer SuperMUC at Leibniz Supercomputing Centre (LRZ, http://www.lrz.de).

During the work, we make use of the following software: {\sc colossus} \citep{Diemer18},
{\sc halotools v0.7} \citep{Hearin2017}, {\sc regressis} \citep{Chaussidon2022}, {\sc handy} \citep{handy}.

\section*{Data Availability}
The halo catalogues and CCMD mocks are publicly available at
\url{https://www.astro.utah.edu/~zhengzheng/data.html}. 
All intermediate data underlying the paper figures/results 
are available on reasonable request.



\bibliographystyle{mnras}
\bibliography{ms} 



\appendix

\section{On the double-peak feature in the group-based CLF of central galaxies}

In this appendix, we investigate the double-peak feature in the CLF of central galaxies from some  group mass bins, as seen in Fig.\ref{fig:CLF_CCMD_groups_SDSS_groups}. 
As the comparisons in this section are restricted to the CCMD groups (i.e., there is no comparison between mocks and actual data), we only considered the redshift and flux-limit cut during the construction of CCMD mock redshift surveys, and we did not incorporate observational effects such as fibre collisions and the SDSS footprint (except in Fig.~\ref{fig:double_peak_meanz}).

\label{sec:clf_double_peak}
\subsection{Why was a prominent double-peak feature in the CLF not seen in previous work?}
\label{subsec:why_no_feature}

\begin{figure*}
\includegraphics[width=\textwidth]{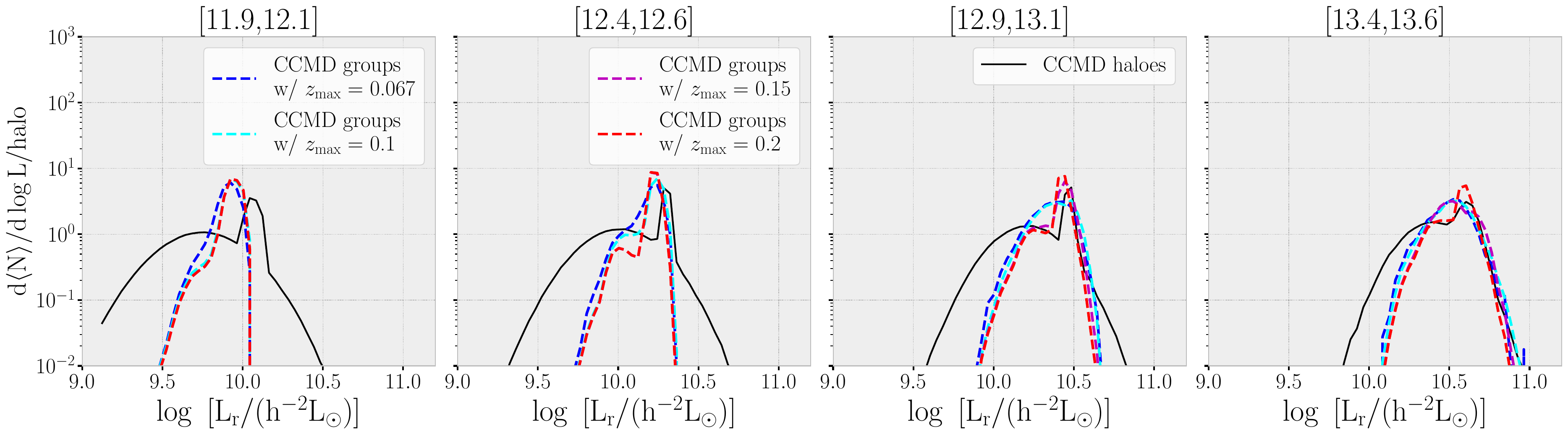}
\caption{
Dependence of the group-based CLF of central galaxies on evaluation volume from the CCMD mock. The evaluation volume to compute the CLF is determined by the redshift cut $z_{\max}$.
The double-peak feature is prominent when evaluated across the entire volume ($z_{\rm max} = 0.2$), but begins to diminish as the volume of evaluation reduces. For comparison, the halo-based CLF of central galaxies is plotted as the solid curve in each panel. The range of halo/group mass in terms of $\logMh$ is indicated at the top of each column.
}
\label{fig:CLF_double_peak}
\end{figure*}

The double-peak feature in some mass bins of the group-based CLF of central galaxies is not as prominent in previous studies. For example, in fig.2 of \citet{Yang08}, the group-based CLF slightly displays a minor hint of the double-peak feature in the central CLF of $12.9 < \logMg < 13.2$ groups. The feature is also barely seen in fig.1 of \citet{Meng2022}.

In Fig.~\ref{fig:CLF_double_peak}, when the group-based CLF is evaluated across the entire volume (i.e. $z_{\rm max} = 0.2$) of the CCMD mock, it shows a distinct double peak feature. However, when the CLF is evaluated in a reduced volume, the double-peak feature begins to diminish, resembling the CLF observed in previous studies.

\begin{figure}
\includegraphics[width=\columnwidth]{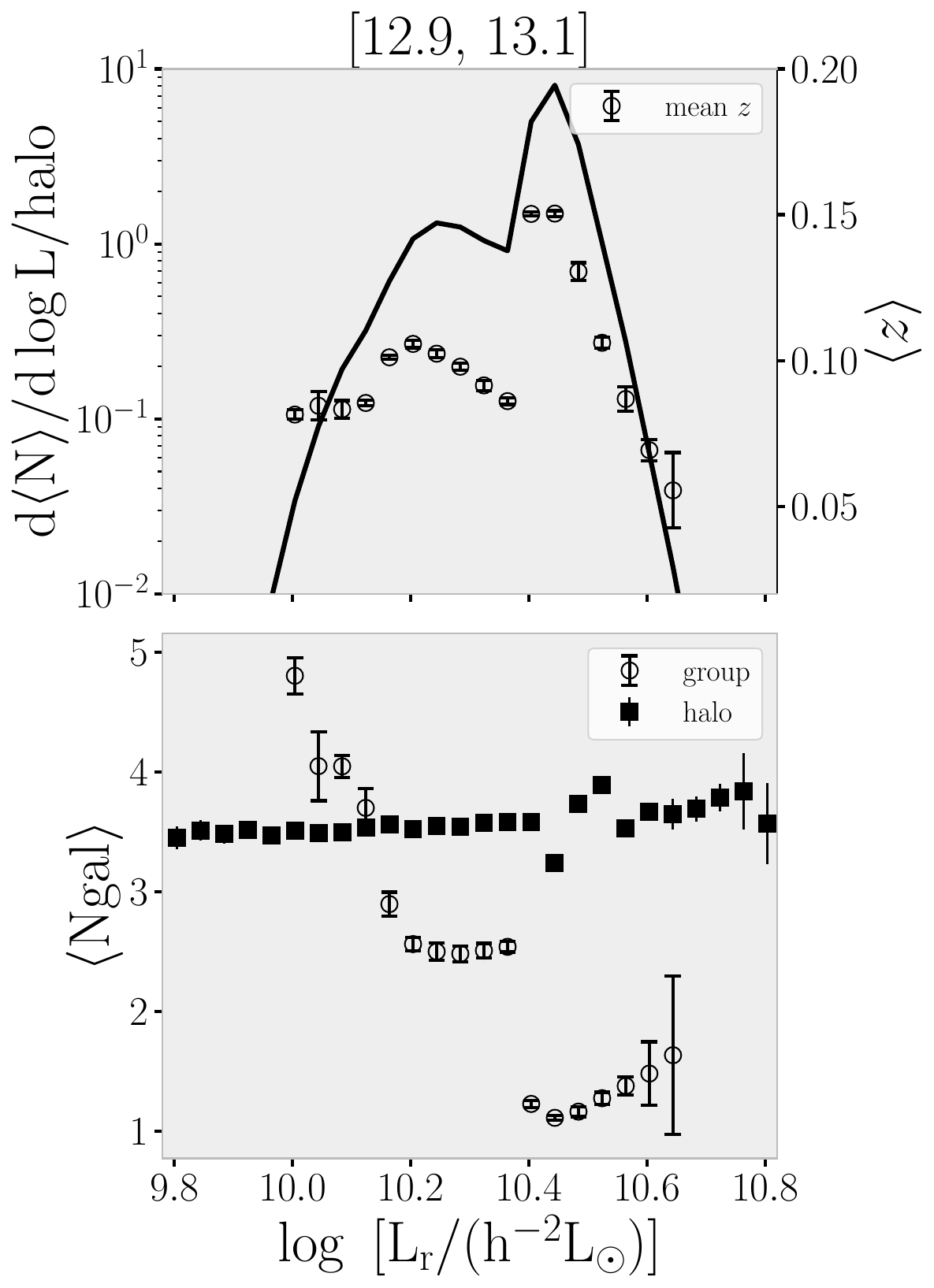}
\caption{
    Group-based CLF of central galaxies at group mass $\logMg\sim 13$ and properties of the groups used to compute the CLF, based on groups identified in the CCMD mock. The solid curve is the CLF from groups in a volume with $z_{\rm max} = 0.2$. The empty circles in the upper panel indicate the mean redshift of central galaxies as a function of luminosity, while circles in the bottom panel represent the mean number of member galaxies in the groups as a function of the central galaxy luminosity. The errorbars on the circles are estimated from groups in CCMD quadrants (Section.~\ref{subsubsec:octant}). 
    To compare, We overplot the mean number of galaxies as a function of central luminosity for haloes (rather than groups) with similar mass, shown as the solid squares. The little fluctuation at $\log~[L_{r}/(\hinvhinvLsun)] \sim 10.5$ is caused by the discontinuity in the scatter of absolute magnitude of the pseudo-blue central galaxy - halo mass relation (top middle panel of Fig.12 in Xu18) at a halo mass with $\logMh \sim 13$. We have verified that this fluctuation will be gone at different halo masses.}
\label{fig:double_peak_meanz}
\end{figure}

We investigate the cause of the disappearance of the more luminous peak by selecting a group mass bin of $12.9<\logMg<13.1$, where the CLF profile changes significantly as the volume decreases.
In Fig.~\ref{fig:double_peak_meanz}, we measure the mean redshift of the central galaxies as a function of luminosity. Surprisingly, there is a peak in the mean redshift distribution, located precisely at the luminosity of the more luminous peak. This appears to explain why the more luminous peak fades away in limited volume. We find that most of these more luminous centrals are at higher redshifts and are the only members of their host groups, with the satellites being too faint to be detected. As a consequence of the group mass assignment (based on the rank of group luminosity), those single-member groups at higher redshifts are assigned a group mass similar to groups at lower redshifts with less luminous centrals but with detectable satellites.

\begin{figure*}
\includegraphics[width=\textwidth]{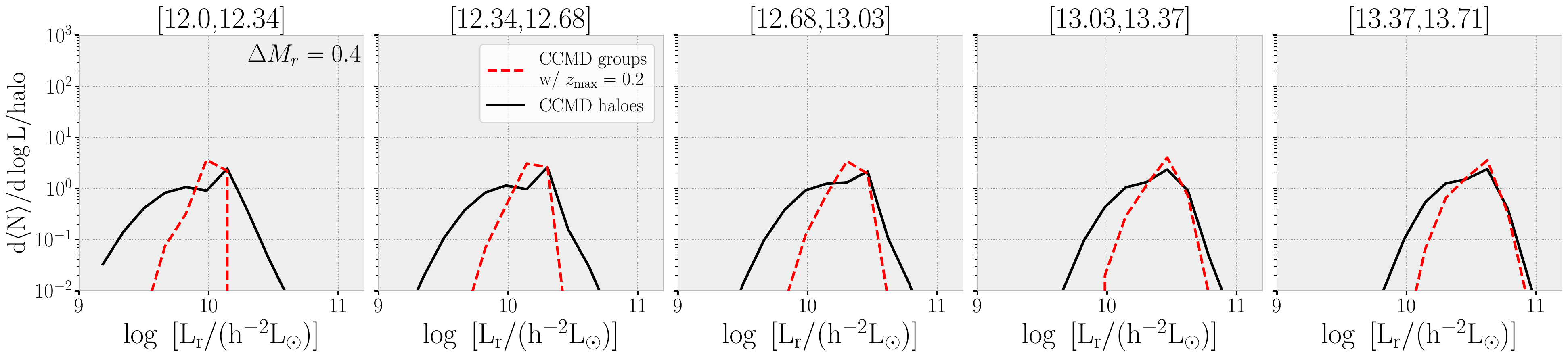}
\caption{
Similar to Fig.~\ref{fig:CLF_double_peak} for $z_{\rm max}=0.2$, but with a larger group mass bin width and a larger luminosity bin width ($\Delta M_r = 0.4$ mag; same as in fig.1 of \citealt{Meng2022}). Compared to the prominent double-peak feature seen in Figure~\ref{fig:CLF_double_peak}, the CLF evaluated with a coarser bin of group mass and luminosity appears to be close to a Gaussian profile, as seen in previous studies.
}
\label{fig:CLF_Meng2022}
\end{figure*}

In addition to the volume effect, both the bin widths of the group mass and the luminosity used to show the CLF would also affect the prominence of the double-peak feature, as demonstrated in Fig.~\ref{fig:CLF_Meng2022}. CLF presented with a wider galaxy luminosity bin from groups of a larger range of group mass would smear out the double-peak feature.

In summary, the double-peak feature in the central galaxy CLF should be present as long as the group mass is determined by the rank of group luminosity. It can be obscured by a variety of factors, such as the volume used to choose galaxy groups for computing the CLF, the group mass bin width, and the luminosity bin width.

\subsection{Does the double-peak feature seen in the group-based central CLF indicate two CCMD components?}
\label{subsec:validate_or_not}

The group-based central CLF exhibits a double-peak profile when measured from the entire volume, which looks like the superposition of a wide Gaussian profile and a narrow Gaussian profile. This feature seems to back up the CCMD modelling results of two orthogonal central components, a pseudo-blue (pseudo-red) population narrow (broad) in luminosity (Xu18).

\begin{figure*}
\includegraphics[width=\textwidth]{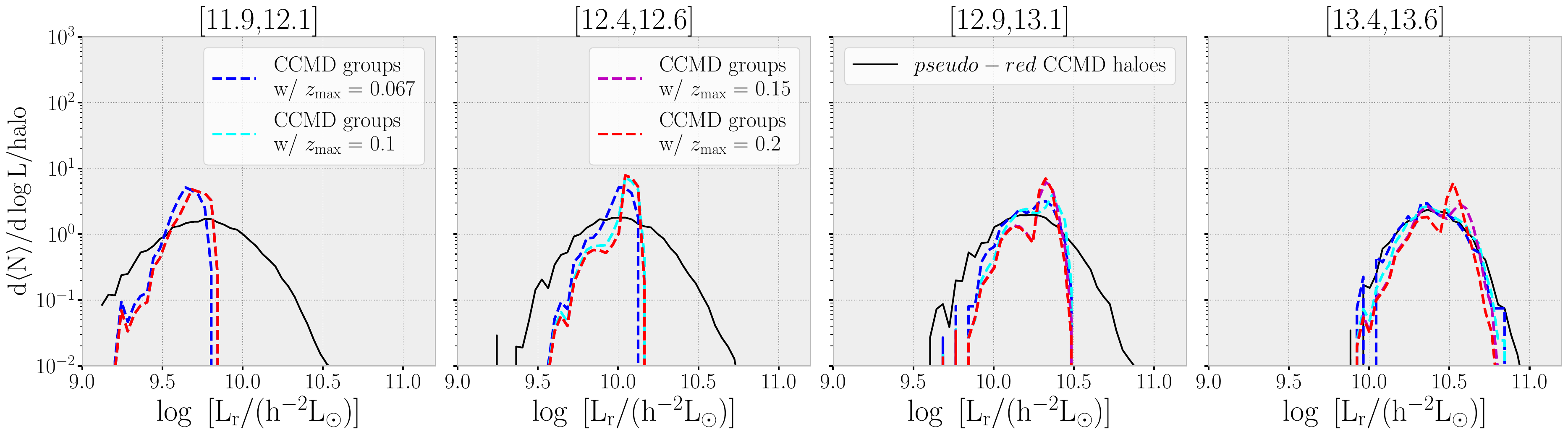}
\caption{
Similar to Fig.~\ref{fig:CLF_double_peak}, but from the CCMD mock with one central galaxy component removed. The pseudo-blue central galaxies are removed from the CCMD mock before applying the group finder. In this modified mock, the shape of the input halo-based CLF for central galaxies is Gaussian (solid curves). However, when estimated in large volume, the group finder produces an artificial double-peak profile in the CLFs (dashed curves for different redshift cuts).
}
\label{fig:CLF_pbcen_removed}
\end{figure*}

However, our test rules out such an explanation. 
In previous section,
we already show that the narrow peak feature in the central CLF is a group mass assignment effect. It should not depend on the exact shape of the central CLF. In Fig.~\ref{fig:CLF_pbcen_removed}, we demonstrate this by applying the group finder to a CCMD mock with only the pseudo-red central component, whose halo-based CLF has a wide Gaussian profile (black solid curve). As we can see from the figure, the measured CLFs from groups across the entire volume display a double-peak feature.

Given the above investigations, we conclude that when using galaxy groups to study galaxy CCMD, additional care should be taken in connecting group-based CCMD to halo-based CCMD and the group finding algorithm needs to be accounted for to interpret the features.

\section{Comparisons with the {\tt Tinker2020} Group Finder}
\label{sec:Tinker2020}

\begin{figure*}
\includegraphics[width=\textwidth]{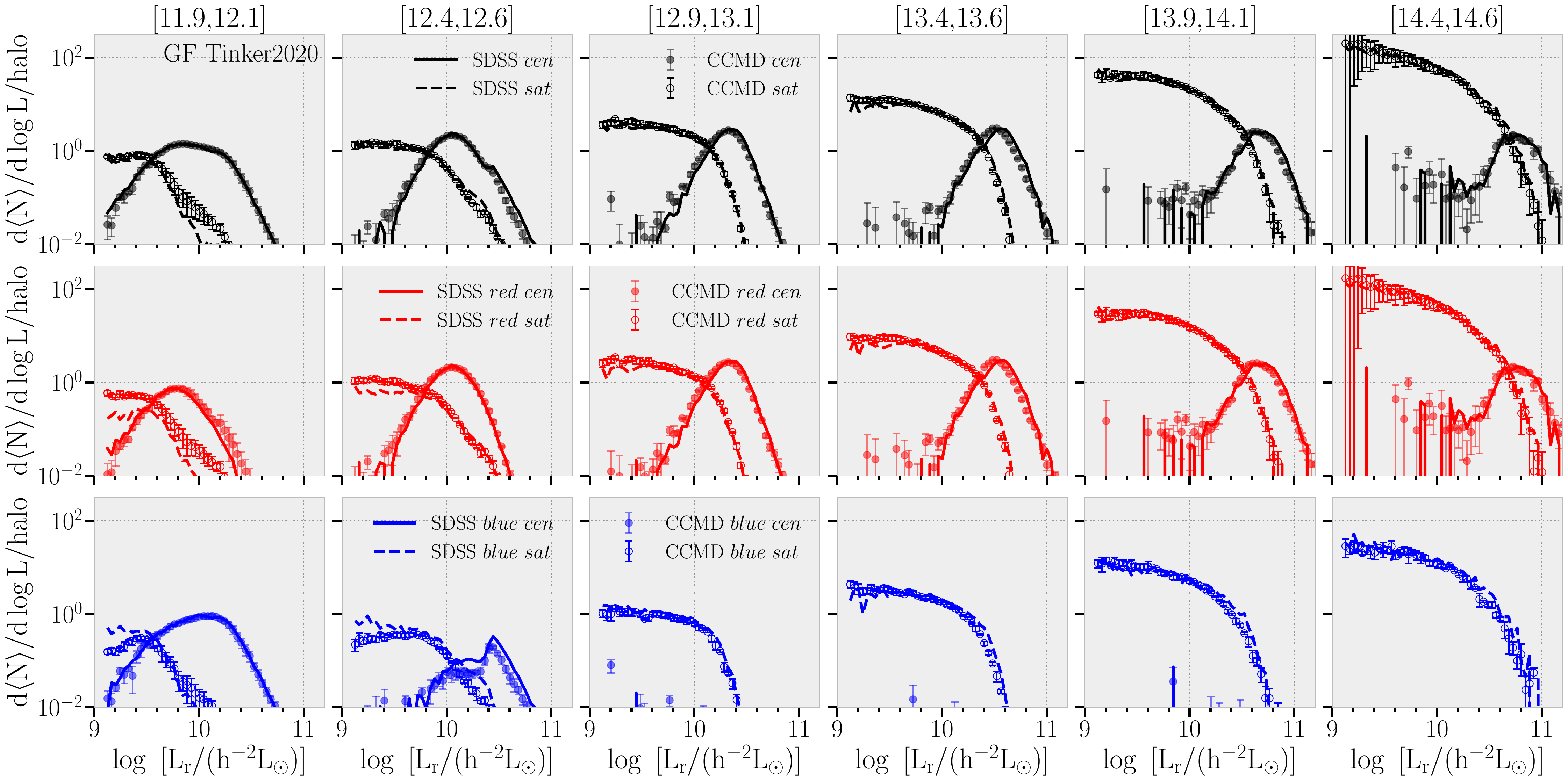}
\caption{
Similar to Fig.~\ref{fig:CLF_CCMD_groups_SDSS_groups}, but the CLFs are from the CCMD groups and the SDSS groups identified by applying the {\tt Tinker2020} group finder to both the CCMD quandrants and SDSS DR7. More information can be found in Section~\ref{sec:Tinker2020}.
}
\label{fig:TinkerGF_CLF_CCMD_groups_SDSS_groups}
\end{figure*}

In the main text, we present the comparison between SDSS groups and CCMD groups based on the $\Yang$ group finder, which suggests that the CCMD inferred from Xu18 provides a good description of the relation between galaxy colour and luminosity and halo mass. For the purpose of the work in this paper, it does not matter which group finder is adopted. Nevertheless, it is useful to make comparisons with another group finder with different performances and systematics. 
In this section, we apply the {\tt Tinker2020} group finder to both SDSS DR7 and the CCMD mocks and compare the results.

The {\tt Tinker2020} group finder \citep{Tinker2020, Tinker2021} was designed to expand the halo-based group-finding algorithm. This extension takes into account the colour-dependent clustering and total satellite luminosity from deep imaging surveys, to account for any potential unknown discrepancies between the galaxy-halo connection for star-forming and quiescent galaxies.
After testing the algorithm against mocks, they implemented a more sophisticated halo mass assignment algorithm. By applying this updated version group finder to the SDSS DR7, they constructed a group catalogue that well reproduces the halo mass estimated by weak lensing analysis for both star-forming and quiescent central galaxies. See full details in \citet{Tinker2021}. 

The {\tt Tinker2020} group finder and its SDSS DR7 group catalogue are publicly available (\url{https://www.galaxygroupfinder.net/}). To make a fair comparison, we apply the {\tt Tinker2020} group finder to the VAGC $bright1$ catalogue, against which the CCMD parameters were tuned. The resulting group catalogue may differ from the public one (which is constructed from $bright34$). Before running the group finder, we apply the same magnitude and redshift cut to the SDSS data (Section~\ref{subsec:sdss}) and with additional observational effects, including the SDSS footprint and fibre collision, 
to the CCMD mocks (Sections~\ref{subsubsec:mock_generate} and ~\ref{subsubsec:octant}), as in the main text.

In addition to galaxy positions, luminosity, and colour, a secondary parameter (such as stellar velocity dispersion, galaxy size, and morphological information) is needed as input information to the {\tt Tinker2020} group finder for more accurate halo mass/membership assignments. Among the list of secondary parameters, they chose the galaxy concentration, which is the ratio between the radius that contains 90 per cent of the galaxy light to the half-light radius. However, the CCMD mock does not have such information for galaxies. As a solution, each CCMD mock galaxy is assigned the concentration of a SDSS galaxy that is closest to it in the colour-magnitude space. 
The code outputs a satellite probability for each galaxy to define central and satellite galaxies, and we use the default threshold, $P_{\rm sat} > 0.5$, to distinguish the two.   

Note that when applying the {\tt Tinker2020} group finder to both the VAGC $bright1$ catalogue and the CCMD mocks, we employ the best-fit free parameters listed in Table 1 of \citet{Tinker2021}. Since our data set is different from the public version (i.e. $bright1$  VS. $bright34$), in principle, we should re-tune these parameters by fitting the colour-dependent clustering and total satellite luminosity present in our data set during the group-finding process. In other words, the configuration we used here might not be optimal in terms of the performance of the finding groups. However, as we emphasise in the main text, the comparisons hold as long as the same group finder is applied to both the data and the mock, regardless of its performance. 
Furthermore, the best-fit parameters presented in \citet{Tinker2021} should not be too far from the optimal ones, given the small difference between these two data sets.

\begin{figure*}
\includegraphics[width=\textwidth]{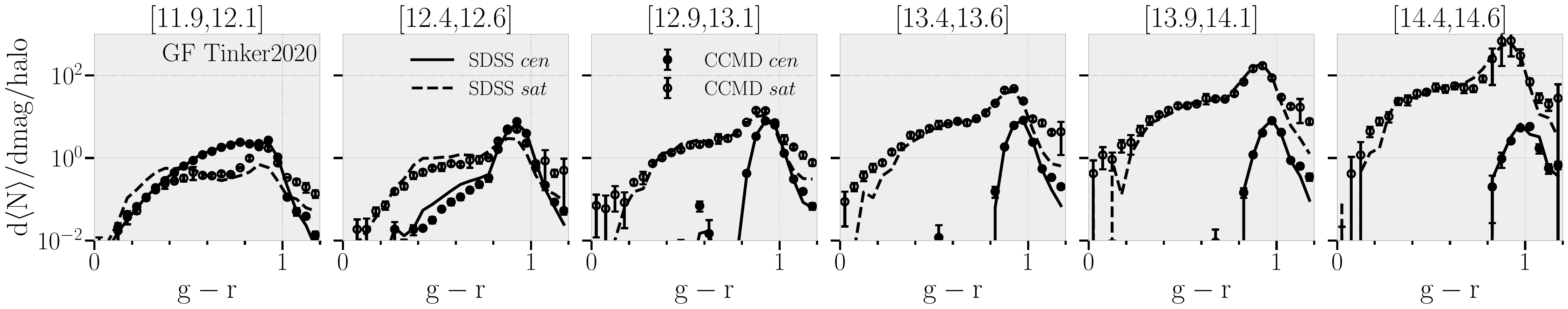}
\caption{
Similar to Fig.~\ref{fig:CCF_CCMD_groups_SDSS_groups}, but the CCFs are from groups identified with the {\tt Tinker2020} group finder.
}
\label{fig:TinkerGF_CCF_CCMD_groups_SDSS_groups}
\end{figure*}

\begin{figure*}
\includegraphics[width=\textwidth]{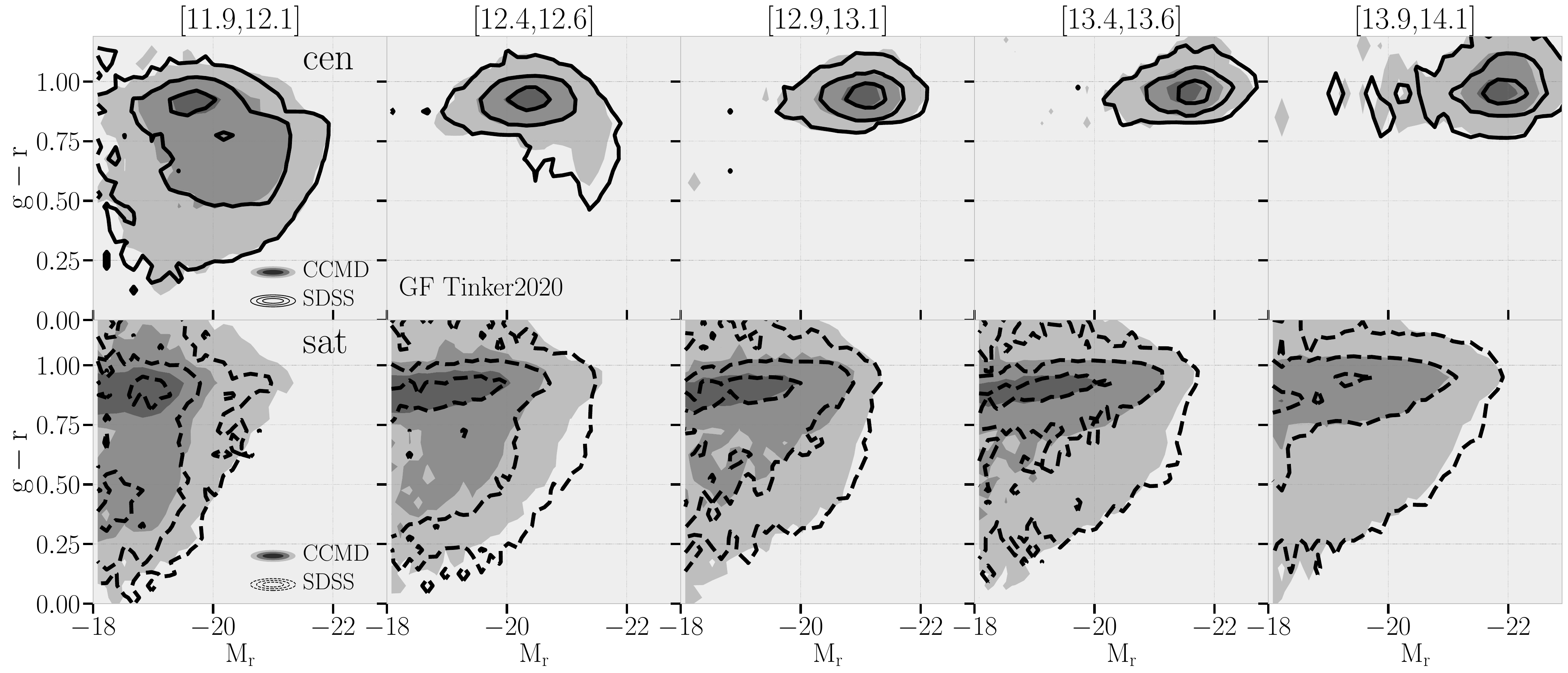}
\caption{
Similar to Fig.~\ref{fig:CCMD_CCMD_groups_SDSS_groups}, but the CCMDs are from groups identified with the {\tt Tinker2020} group finder.
}
\label{fig:TinkerGF_CCMD_CCMD_groups_SDSS_groups}
\end{figure*}

Comparisons of CLF, CCF, and CCMD between the CCMD groups and SDSS groups from the {\tt Tinker2020} group finder are shown in Fig.~\ref{fig:TinkerGF_CLF_CCMD_groups_SDSS_groups}, Fig.~\ref{fig:TinkerGF_CCF_CCMD_groups_SDSS_groups}, and Fig.~\ref{fig:TinkerGF_CCMD_CCMD_groups_SDSS_groups}, respectively. 
Overall, the trends seen in the results are similar to those in Fig.~\ref{fig:CLF_CCMD_groups_SDSS_groups}, Fig.~\ref{fig:CCF_CCMD_groups_SDSS_groups}, and Fig.~\ref{fig:CCMD_CCMD_groups_SDSS_groups}, based on the {\tt Yang et al.} group finder. This again suggests that the CCMD in Xu18 provides a good description of the galaxy-halo relation in terms of galaxy colour/luminosity and halo mass. However, in detail, the group-based CLF, CCF, and CCMD depend on the group finder. For example, the {\tt Tinker2020} group finder almost eliminates double-peak features in the CLF of the CCMD mocks, while the $\Yang$ group finder would create an artificial peak feature, especially when single-member systems at relatively higher redshift are included (Fig.~\ref{fig:CLF_pbcen_removed}). In massive groups identified by the {\tt Tinker2020} group finder, blue central galaxies are removed. 
The reason for the absence of double-peak feature in  {\tt Tinker2020} central CLF may link to the group mass assignment in the {\tt Tinker2020} group finder, which is jointly determined by total luminosity, central galaxy's colour, and the correlation between total luminosity of faint satellites from deep imaging survey and halo mass.

Comparing the group galaxy statistics from different group finders and their connection to the ground truth of galaxies in haloes, based on CCMD mocks can reveal common trends and interesting differences, which will be further explored in a separate work.

\section{Joint prediction of galaxy colour and absolute magnitude from a halo catalogue}

In addition to the provided halo and CCMD mock catalogues, we present the best-fit CCMD parameters here to ensure reproducibility and to facilitate the creation of colourful galaxy mocks using alternative halo catalogues of the reader's choice.

The CCMD formalism has been briefly introduced in the section~\ref{sec:intro} and more details can be found in Xu18. In this appendix, we only list the necessary gradients for CCMD mock construction from halo catalogues.
Central galaxies are assumed to reside at the potential
minimum of their host haloes and inherit their bulk velocities. 
For the satellite galaxies, we assign them the positions and velocities
of randomly chosen dark-matter particles.
We note that the colour ($g - r$) and absolute magnitude (in $r$ band, setting $h = 1$) assigned in the CCMD mocks are of Petrosian magnitude and have been $K$-corrected to $z \sim 0.1$. We have also assumed that the colour and luminosity are determined only by (viral) halo mass, i.e., there is no assmebly bias in the CCMD mocks. 

\subsection{CCMD of central galaxies}
To be more self-contained, we briefly introduce the formalism of CCMD central galaxies in this section. For more details, see Section 3.1 in Xu18.

At a given halo mass, the colour and absolute magnitude distribution of central galaxies are assumed to be composed by two populations, pseudo-red and pseudo-blue central galaxies, both of which are assumed to follow a 2D Gaussian distribution in colour and absolute magnitude plane. 
For convenience of presentation, we refer to the magnitude direction 
as the $x$ direction and the colour direction as the $y$ direction. The differential colour-magnitude distribution of each pseudo-colour central
galaxy component can be written as
\begin{equation}
\frac{{\rm d}^2 \langle N_{\textrm{cen, comp}} \rangle}{\rm dx\, dy} =
f_{\rm comp}\frac{1}{2\pi\sigma_{\rm x}\sigma_{\rm y}
\sqrt{1-\rho^2}}\exp\left[-\frac{Z^2}{2(1-\rho^2)}\right],
\end{equation}
with
\begin{equation}
Z^2 =  \frac{(x-\mu_{\rm x})^2}{\sigma^2_{\rm x}}+\frac{(y-\mu_{\rm y})^2}
 {\sigma^2_{\rm y}} - \frac{2\rho(x-\mu_{\rm x})(y-\mu_{\rm y})}
 {\sigma_{\rm x}\sigma_{\rm y}}
\end{equation}
and
\begin{equation}
\rho=\frac{Cov(x,y)}{\sigma_{\rm x}\sigma_{\rm y}}.
\end{equation}
Here `comp' is `p-red' (`p-blue') for the pseudo-red (pseudo-blue) population, $\mu_{\rm x}$ is the mean magnitude, $\mu_{\rm y}$ is the mean colour, $\sigma_{\rm x}$ and $\sigma_{\rm y}$ are the standard deviations in magnitude and colour, and $\rho$ ($Cov(x,y)$) is  the coefficient of the correlation (covariance) between magnitude and colour.

We now parameterize the halo mass dependence for the above parameters.For each pseudo-colour central galaxy population, we first parameterize the halo-mass dependent mean and scatter in magnitude. 
\begin{equation}
\label{eqn:mux}
\mu_{\rm x} = x_{\rm t}-2.5\alpha_{\rm M}\log
\left( \frac{\Mh}{M_{\rm t}} \right) - 1.086
\left(-\frac{M_{\rm t}}{\Mh}+1\right),
\end{equation}
with $\mu_{\rm x}$ the mean absolute magnitude of the central galaxy population 
in haloes of mass $M_{\rm h}$ and $x_{\rm t}$ that in haloes of transition mass 
$M_{\rm t}$.
For the scatter in the magnitude of central galaxies of each population, 
we parameterize it to follow a linear relation with halo mass, motivated
by \citet{Yang08} (see their fig.4),
\begin{equation}
\label{eqn:sigx}
\sigma_{\rm x} = a_{\rm 0,\sigma_x} 
+ a_{\rm 1,\sigma_x} (\log \Mh - \log M_{\rm nl}),
\end{equation}
with two free parameters, $a_{\rm 0,\sigma_x}$ and $a_{\rm 1,\sigma_x}$. 
The quantity $M_{\rm nl}$ is the $z=0$ nonlinear mass for collapse, and for
our adopted cosmology $\log [M_{\rm nl}/(\hinvMsun)]=12.35$.

For the mean colour $\mu_{\rm y}$ and colour scatter $\sigma_{\rm y}$ of 
central galaxies, we parameterize both $\mu_{\rm y}$ 
and $\sigma_{\rm y}$ as a function of the mean magnitude $\mu_{\rm x}$, which 
is then linked to the halo mass. we parameterize them as a combination of a linear function and a hyperbolic tangent function,
\begin{equation}
\label{eqn:muy}
\mu_{\rm y} = a_{\rm 0,y} + a_{\rm 1,y}(\mu_{\rm x}-\mu_{\ast})
+q_{\rm 0,y}\textrm{tanh}\left(\frac{\mu_{\rm x}-q_{\rm 1,y}}
{q_{\rm 2,y}}\right)
\end{equation}
and
\begin{equation}
\label{eqn:sigy}
\sigma_{\rm y} = 
a_{\rm 0,\sigma_y} + a_{\rm 1,\sigma_y}(\mu_{\rm x}-\mu_\ast) 
+q_{\rm 0,\sigma_y} \rm{tanh}\left(\frac{\mu_x-q_{\rm 1,\sigma_y}}{q_{\rm 2,\sigma_y}}\right),
\end{equation}
where $a$'s and $q$'s are free parameters and $\mu_*=-20.44$ corresponds to 
the $r$-band characteristic luminosity in the local galaxy LF.

Since the correlation $\rho$ between central galaxy luminosity and colour at
fixed halo mass is not well constrained in the literature, a simple  
linear relation is proposed to describe its halo mass dependence,
\begin{equation}
\label{eqn:rho}
\rho = a_{\textrm{0,$\rho$}} + a_\textrm{1,$\rho$}(\log M_h - \log M_{\rm nl}), 
\end{equation}
with $a_\textrm{1,$\rho$}$ and $a_{\textrm{0,$\rho$}}$ as the two 
free parameters.
 
For the fraction of pseudo-red central galaxies, the galaxy CMD suggests that it increases with increasing galaxy luminosity, and hence halo mass. We 
parameterize it with a ramp-like function,
\begin{equation}
\label{eqn:fred}
f_{\textrm{p-red}} = \frac{1}{2}a_{\textrm{0,f}}\left[ 1 +
\textrm{erf}\left(\frac{\log 
M_h-a_{\textrm{2,f}}}{a_\textrm{1,f}}\right)\right],
\end{equation}
where ${\rm erf}$ is the error function.

We list the best-fit values and their uncertainties for CCMD central galaxies in Table ~\ref{tab:cenpara}.

\begin{table}
\caption{The colour-magnitude distribution of central galaxies is described as a combination of contributions from two galaxy populations, the pseudo-red (`p-r' in the table) and pseudo-blue (`p-b') central galaxies. At fixed halo mass, the distribution of each population is modelled as a 2D Gaussian. The halo-mass
dependences of the quantities in each 2D Gaussian function are described
with 17 parameters: 3 for the mean magnitude ($x_{\rm t}$, $M_{\rm t}$, and
$\alpha_{\rm M}$; Eq.~\ref{eqn:mux}), 2 for the scatter in magnitude 
($a_{0,\sigma_{\rm x}}$ and $a_{1,\sigma_{\rm x}}$; Eq.~\ref{eqn:sigx}),
5 for the mean colour ($a_{\rm 0,y}$, $a_{\rm 1,y}$, $q_{\rm 0,y}$, 
$q_{\rm 1,y}$, and $q_{\rm 2,y}$; Eq.~\ref{eqn:muy}), 5 for the scatter in 
colour ($a_{\rm 0,\sigma_y}$, $a_{\rm 1,\sigma_y}$, $q_{\rm 0,\sigma_y}$, 
$q_{\rm 1,\sigma_y}$, and $q_{\rm 2,\sigma_y}$; Eq.~\ref{eqn:sigy}), and 2 
for the colour-magnitude correlation ($a_{0,\rho}$ and $a_{1,\rho}$;
Eq.~\ref{eqn:rho}). Therefore, the two 2D Gaussian functions for the two 
central galaxy populations have 34 parameters. There are also 3
parameters ($a_{\rm 0,f}$, $a_{\rm 1,f}$, and $a_{\rm 2,f}$; Eq.~\ref{eqn:fred})
in the fraction of the pseudo-red central galaxies, which specifies the 
relative normalisation of the two Gaussian functions. In total, we use 37
parameters to describe the CCMD of central galaxies.}
    \centering
    \begin{tabular}{ccccc}
        \hline
        \hline
         \multirow{2}*{} & best-fit & median & 16\% & 84\% \\
           & `p-b'/`p-r' & `p-b'/`p-r' & `p-b'/`p-r' & `p-b'/`p-r' \\
         \hline
         \multicolumn{5}{c}{mean magnitude, Eq.~\ref{eqn:mux}}  \\
         \hline
         $x_{\rm t}$ & 11.44/11.42 & 11.44/11.50 & 11.43/11.44 & 11.45/11.69 \\
         $M_{\rm t}$ & -19.30/-18.38 & -19.27/-18.53 & -19.33/-18.93 &  -19.23/-18.40  \\
         $\alpha_{\rm M}$ & 0.28/0.35 & 0.29/0.34 & 0.28/0.28 & 0.29/0.35  \\
         \hline
         \multicolumn{5}{c}{scatter in magnitude, Eq.~\ref{eqn:sigx} }\\
         \hline
         $a_{0,\sigma_{\rm x}}$ & -0.15/0.56 & -0.19/0.54 & -0.25/0.49 &  -0.14/0.57   \\
         $a_{1,\sigma_{\rm x}}$ & 0.26/-0.12 & 0.30/-0.09 & 0.23/-0.12 &  0.37/-0.03   \\      
        \hline
        \multicolumn{5}{c}{mean colour, Eq.~\ref{eqn:muy}}\\
        \hline
         $a_{\rm 0,y}$ & 0.88/0.77 & 0.87/0.79 & 0.86/0.77 & 0.88/0.87   \\
         $a_{\rm 1,y}$ & -0.05/-0.04 & -0.04/-0.05 & -0.05/-0.05 & -0.04/-0.04  \\     
         $q_{\rm 0,y}$ & -0.40/-0.18 & -0.40/-0.16 & -0.41/-0.18 & -0.37/-0.07    \\
         $q_{\rm 1,y}$ &  -22.49/-18.86 & -22.39/-18.78 & -22.47/-18.91 & -22.31/-18.53    \\   
         $q_{\rm 2,y}$ & 2.67/0.59 & 2.87/0.62 & 2.63/0.54 & 2.96/0.81   \\
        \hline
        \multicolumn{5}{c}{scatter in colour, Eq.~\ref{eqn:sigy}}\\
        \hline
         $a_{\rm 0,\sigma_y}$ & 0.22/0.04 & 0.22/0.05 & 0.18/0.03 & 0.28/0.06 \\
         $a_{\rm 1,\sigma_y}$ & -0.10/0.06 & -0.09/0.05 & -0.10/0.04 & -0.07/0.08  \\     
         $q_{\rm 0,\sigma_y}$ & 0.21/-0.10 & 0.21/-0.10 & 0.11/-0.11 & 0.29/-0.09 \\
         $q_{\rm 1,\sigma_y}$ & -19.49/-21.09 & -19.50/-21.38 & -19.96/-21.78 & -18.71/-20.86  \\   
         $q_{\rm 2,\sigma_y}$ & 2.21/1.85 & 2.21/1.82 & 1.39/1.39 & 2.99/2.54  \\
        \hline
        \multicolumn{5}{c}{colour-magnitude correlation, Eq.~\ref{eqn:rho}}\\
        \hline
         $a_{0,\rho}$ & -0.48/0.69 & -0.29/0.69 & -0.40/0.55 & -0.15/0.82  \\
         $a_{1,\rho}$ & 0.20/-0.08 & 0.06/-0.03 & -0.08/-0.10 & 0.17/0.02 \\ 
        \hline
        \multicolumn{5}{c}{normalisation of the two Gaussian functions, Eq.~\ref{eqn:fred}}  \\
        \multicolumn{5}{c}{Note: no `p-b'/`p-r' distinction in the following parameters.}  \\
        \hline
         $a_{\rm 0,f}$ &  0.66 & 0.65 & 0.62 & 0.67\\
         $a_{\rm 1,f}$ &  0.42 & 0.39 & 0.35 & 0.46\\  
         $a_{\rm 2,f}$ &  11.25 & 11.32 & 11.25 & 11.39\\
         \hline
    \end{tabular}
    
    \label{tab:cenpara}
\end{table}

\subsection{CCMD of Satellite Galaxies}
\label{subsec:satpara}

For satellite galaxies, we also divide their CCMD into contributions from two 
populations, namely the pseudo-blue and pseudo-red satellite galaxies. The satellite CCMD at fixed halo mass for a pseudo-colour population reads
\begin{multline}
\label{eqn:satCCMD}
\frac{{\rm d}^2 \langle N_{\textrm{sat,comp}} \rangle}{\rm dx\, dy} 
 =
 \frac{1 }{\sqrt{2\pi}\sigma_{\textrm{y,sat}}} \exp\left[- \frac{
 (y-\mu_{\textrm{y,sat}})^2}{2\sigma^2_{\textrm{y,sat}}} \right]  \\
 \times 0.4\phi_{\rm s}
10^{-0.4(\alpha_{\rm s}+1)(x-x_{\rm s}^\ast)}  
\exp \left[ -10^{-0.8(x-x_{\rm s}^\ast)}  \right], 
\end{multline}
where $x_{\rm s}$ is the absolute magnitude corresponding to $L_{\rm s}^\ast$,
the Gaussian function represents the colour distribution (see below), and
similar to the case with central galaxies `comp' here can be either `p-blue'
or `p-red'.

Each quantity in the right-hand side of Eq.~\ref{eqn:satCCMD} varies with halo
mass. Motivated by previous research, we adopt a quadratic function to describe the halo-mass dependence of both the normalisation $\phi_{\rm s}$
and the faint-end slope $\alpha_{\rm s}$.
\begin{multline}
\label{eqn:phis}
\log \phi_{\rm s} = a_{\textrm{0,$\phi$}} + a_{\textrm{1,$\phi$}} (\log \Mh-\log \Mnl) \\ 
+ a_{\textrm{2,$\phi$}} (\log \Mh -\log \Mnl)^2
\end{multline}
and
\begin{multline}
\label{eqn:alphas}
\alpha_{\rm s} = a_{\textrm{0,$\alpha$}} + a_{\textrm{1,$\alpha$}}(\log  \Mh - \log \Mnl)\\
 + a_{\textrm{2,$\alpha$}} (\log  \Mh - \log \Mnl)^2,
\end{multline}
where the $a$'s are free parameters.

\cite{Yang08} suggest that there is a luminosity gap between the median 
luminosity of central galaxy $L_\textrm{cen}$ and the characteristic
luminosity $L_{\rm s}^\ast$ of satellite galaxies independent of halo mass. 
In terms of absolute magnitude, we have
\begin{equation}
\label{eqn:xs}
x_{\rm s}^\ast = \mu_x + \Delta_{\rm sc},
\end{equation}
where $\Delta_{\rm sc}$ (independent of halo mass) is the luminosity gap in 
absolute magnitude. This $\Delta_{\rm sc}$ parameter and the halo-mass dependent
$\mu_{\rm x}$ (Eq.~\ref{eqn:mux}) give our parameterization of the 
dependence of $x_{\rm s}^\ast$ on halo mass.

We parameterize the mean and standard deviation of the satellite colour as a linear function with absolute magnitude,
\begin{equation}
\label{eqn:muys}
\mu_{\textrm{y,sat}} = a_{\textrm{0,y,sat}} + a_{\textrm{1,y,sat}}(x-\mu_{\ast})
\end{equation}
and
\begin{equation}
\label{eqn:sigys}
\sigma_{\textrm{y,sat}} = a_{\textrm{0, $\sigma_{\textrm{y,sat}}$}} + a_{\textrm{1, $\sigma_{\textrm{y,sat}}$}}(x-\mu_{\ast}),
\end{equation}
where the $a$'s are parameters to link the satellite colour distribution with
the magnitude and $\mu_\ast=-20.44$ corresponds to the characteristic luminosity
in the $r$-band LF of local galaxies.

The best-fit values and their uncertainties for CCMD satellite galaxies are listed in Table ~\ref{tab:satpara}.

\begin{table}
\caption{In the parameterization of the satellite CCMD, we introduce 
parameters related to the CLF and satellite colour distribution. For each
pseudo-colour satellite population, there are 7 parameters to describe the
CLF as a function of halo mass: 3 for $\phi_{\rm s}$ ($a_{0,\phi}$,
$a_{1,\phi}$, and $a_{2,\phi}$; Eq.~\ref{eqn:phis}), 3 for $\alpha_{\rm s}$
($a_{0,\alpha}$, $a_{1,\alpha}$, and $a_{2,\alpha}$; Eq.~\ref{eqn:alphas},
and 1 for $x_{\rm s}$ ($\Delta_{\rm sc}$; Eq.~\ref{eqn:xs}). There are 4 
parameters to characterise the satellite colour distribution: 2 for the
mean $\mu_{\rm y,sat}$ ($a_{\rm 0,y,sat}$ and $a_{\rm 1,y,sat}$;
Eq.~\ref{eqn:muys}) and 2 for the scatter $\sigma_{\rm y,sat}$ 
($a_{\rm 0,\sigma_{y,sat}}$ and $a_{\rm 1,\sigma_{y,sat}}$; Eq.~\ref{eqn:sigys}).
Overall, we end up with 22 free parameters to describe the CCMD of 
the satellite population.}
    \centering
    \begin{tabular}{ccccc}
        \hline
        \hline
         \multirow{2}*{} & best-fit & median & 16\% & 84\% \\
           & `p-b'/`p-r' & `p-b'/`p-r' & `p-b'/`p-r' & `p-b'/`p-r' \\
         \hline
         \multicolumn{5}{c}{$\phi_{\rm s}$ of CLF, Eq.~\ref{eqn:phis}}  \\
         \hline
         $a_{0,\phi}$ & -0.12/0.09 & -0.11/0.03 & -0.17/-0.18 & -0.05/0.20 \\
         $a_{1,\phi}$ &  0.87/0.75 & 0.80/0.72 & 0.67/0.57 & 0.92/0.98 \\
         $a_{2,\phi}$ &  -0.23/-0.00 & -0.23/0.02 & -0.29/-0.05 & -0.16/0.06 \\
         \hline
         \multicolumn{5}{c}{$\alpha_{\rm s}$ of CLF, Eq.~\ref{eqn:alphas} }\\
         \hline
         $a_{0,\alpha}$ & -0.01/-0.60 & -0.18/0.56 & -0.46/-0.41 & -0.05/1.45   \\
         $a_{1,\alpha}$ & -1.47/0.62 & -1.08/-0.74 & -1.34/-1.82 & -0.74/0.31    \\      
         $a_{2,\alpha}$ & 0.36/-0.34 & 0.19/0.00 & 0.07/-0.25 & 0.30/0.28    \\
        \hline
        \multicolumn{5}{c}{$x_{\rm s}$ of CLF, Eq.~\ref{eqn:xs}}\\
        \hline
         $\Delta_{\rm sc}$ & 0.64/1.43 & 0.58/1.37 & 0.52/1.17 & 0.63/1.45  \\
        \hline
        \multicolumn{5}{c}{mean colour $\mu_{\rm y,sat}$, Eq.~\ref{eqn:muys}}\\
        \hline
         $a_{\rm 0,y,sat}$ & 0.89/0.94 & 0.88/0.94 & 0.88/0.94 & 0.89/0.95  \\
         $a_{\rm 1,y,sat}$ & 0.05/-0.03 & -0.06/-0.03 & -0.06/-0.03 &  -0.05/-0.02 \\     
        \hline        
        \multicolumn{5}{c}{scatter in colour $\sigma_{\rm y,sat}$, Eq.~\ref{eqn:sigys}}\\
        \hline
        $a_{\rm 0,\sigma_{y,sat}}$ & 0.12/0.03 & 0.11/0.03 & 0.11/0.03 &  0.12/0.03   \\
        $a_{\rm 1,\sigma_{y,sat}}$ &  0.08/0.00 & 0.08/0.00 & 0.07/0.00 & 0.08/0.01  \\     
         \hline
    \end{tabular}
    
    \label{tab:satpara}
\end{table}


\bsp	
\label{lastpage}
\end{document}